\documentclass[twocolumn,showpacs,prc,superscriptaddress,proprietresses]{revtex4}
\usepackage[dvips]{graphicx}
\usepackage{amsmath,amssymb}
\usepackage{dcolumn}
\begin{document}
\title{Nonextensive hydrodynamics for relativistic heavy-ion collisions}
\author{T. Osada}
\email{osada@ph.ns.musashi-tech.ac.jp}
\affiliation{Theoretical Physics Lab.,~Faculty of Knowledge Engineering,
Musashi Institute of Technology, Setagaya-ku, Tokyo 158-8557, Japan}
\author{G. Wilk}
\email{wilk@fuw.edu.pl}
\affiliation{The Andrzej So{\l}tan Institute for Nuclear Studies, Ho\.{z}a 69, 00681,
Warsaw, Poland}

\date{\today}

\begin{abstract}
The nonextensive one-dimensional version of a hydrodynamical model
for multiparticle production processes is proposed and discussed.
It is based on nonextensive statistics assumed in the form
proposed by Tsallis and characterized by a nonextensivity
parameter $q$. In this formulation the parameter $q$ describes
some specific form of local equilibrium which is characteristic
for the nonextensive thermodynamics and which replaces the local
thermal equilibrium assumption of the usual hydrodynamical models.
We argue that there is correspondence between the perfect
nonextensive hydrodynamics and the usual dissipative
hydrodynamics. It leads to simple expression for dissipative
entropy current and allows for predictions for the ratio of bulk
and shear viscosities to entropy density, $\zeta/s$ and $\eta/s$,
to be made.
\end{abstract}


\maketitle

\section{\label{sec:I}Introduction}
Multiparticle production experiments are our main source of
information on multiparticle production processes in which the
initial kinetic energy of two projectiles is to a large extent
converted into a multitude of observed secondaries. This is
especially true in the case of heavy ion collisions in which one
expects the formation of a new hadronic state of matter, the Quark
Gluon Plasma (QGP) \cite{QGP}. Such processes call for some form
of statistical approach, which is usually based on the
Boltzmann-Gibbs (BG) statistics. On the other hand, in the case of
a multiparticle production process, conditions leading to BG
statistics are satisfied only approximately at best. This is
because, among other things, hadronizing systems experience strong
intrinsic fluctuations and long range correlations
\cite{fluct,fluctl1}, which can be interpreted as signals of some
dynamical, nonequilibrium effects showing up\cite{Foot1}. 
It is
therefore difficult to expect the occurrence of the usual (local)
thermal equilibrium, instead one has some kind of stationary
state. It turns out that these phenomena can be incorporated, at
least to some extent and without going into deeper dynamical
considerations concerning the sources of such fluctuations, in the
formalism of the nonextensive statistics (which we shall apply
here in the manner proposed by Tsallis, see \cite{Tsallis}) in the
form of a more general equilibrium summarily described by a single
parameter $q$ \cite{Biro,BiroK,BiroPurcsel,SR}. This parameter
characterizes the corresponding Tsallis entropy, $S_q$, which in
such approach replaces the usual BG entropy to which it converges
for $q\rightarrow 1$. Because such systems are in general
nonextensive, the parameter $q$ is usually called a {\it
nonextensivity parameter}.

Such approach has been successfully applied to multiparticle
production processes, both by using nonextensive versions of the
respective distribution functions
\cite{E1a,E1b,E2,Kq,QlQt,E3,OUWW}, or by deriving such
distributions from the appropriately modified nonextensive version
of the Boltzmann transport equation
\cite{Biro,BiroK,SR,Lavagno,Lav-appl}. In both cases, this amounts
to replacing the usual exponential factors by their
$q$-exponential equivalents,
\begin{eqnarray}
&& P_{\rm BG}(E) = \exp( - E/T) ~ \Longrightarrow ~  P_q(E) = \exp_q(-E/T) \nonumber \\
&& \hspace*{18mm}= [ 1 - (1-q)\, E/T ] ^{1/(1-q)}. \label{q-q=1}
\end{eqnarray}
Notice that $P_{\rm BG}(E) = P_{q=1}(E)$. In all these
applications one finds that $q > 1$. It represents effect of some
intrinsic fluctuations existing in the hadronizing system and
revealing themselves as fluctuations of its temperature or of the
mean multiplicity of secondaries 
\footnote{
                 The situation in which $q<1$ was  analyzed in
                  \cite{NCql1}. As discussed in \cite{fluctl1},
                  interpretation in terms of fluctuations
                  is not clear now. Instead it was shown that in this case the first
                  role of the parameter $q$ is to restrict the allowed phase space
                  and $q<1$ reflects the fact that only a fraction $K$ (called
                  {\it inelasticity}) of the initially available energy is used for
                  the production of secondaries, the remaining $1-K$ part is to be found
                  in the so called leading particles. As a result one gets $\exp_{q<1}(-X)$
                  distribution with $X$ limited to the $X > 1/(1-q)$ region only.
}.
Generically one
has that \cite{fluct} 
\begin{equation}
q = 1 + \frac{<(1/T)^2>}{<(1/T)>^2}, \label{eq:q_fluct}
\end{equation}
where, depending on the kind of process considered, $T$ can be
replaced by some other variable \cite{Kq,QlQt,E3}. Because
different observables are sensitive to different kinds of
fluctuations, it is natural to expect that they are described by
different values of the parameter $q$ \cite{QlQt}. For example,
single particle, one-dimensional distributions in longitudinal
phase-space, like $dN/dy$, are most sensitive to fluctuations of
the mean multiplicity $\langle n\rangle$ of secondaries \cite{Kq}
and are described by the nonextensivity parameter $q=q_{\rm L}$ of
the order of $q_{\rm L} - 1\sim 0.1 \div 0.2$ 
\footnote{
Actually, these are precisely the same fluctuations that lead to the
                Negative Binomial form of the observed multiplicity distributions,
                $P(n;\langle n\rangle;k)$,
                with its characteristic parameter $k$ given by $k = 1/(q_{\rm L}-1)$ \cite{Kq}
                (in this case one can also speak about fluctuations
                in the so called {\it partition temperature} $T_{pt} = E/\langle n\rangle$,
                see \cite{FluctWW}.
}. 
However, distributions in transverse momenta, $dN/dp_{\rm T}$,
which are believed to probe the local thermal equilibrium of the
hadronizing system (assuming that such a phenomenon indeed occurs)
and serve as a source of information on the temperature $T$ of the
hadronizing system, are very sensitive to fluctuations of this
temperature represented by the nonextensivity parameter $q=q_{\rm
T}$ \cite{E3}: changes as small as $q_{\rm T} - 1 \sim 0.01 \div
0.05$ already substantially affect the resultant $p_{\rm T}$
spectra 
\footnote{
                 In this case $q$ measures fluctuation of the temperature $T$ as given by the
                 specific heat parameter $C$ and $q_{\rm T}-1 = C$
                 \cite{E3,fluct}.  As such it should be inversely proportional to the volume of
                 the interaction region, this effect is indeed observed \cite{E3}.
}
Notice that data indicate that $q_{\rm
L}-1\gg q_{\rm T}-1$, i.e., fluctuations governed by $q_{\rm L}$
are dominant and therefore for the whole system $q \sim q_{\rm L}$
\cite{QlQt}.

Among statistical approaches to the multiparticle production
processes a specially important role is played by hydrodynamical
models \cite{hydro-review}, which so far are all based on the BG
statistics. The existing general nonextensive version of fluid
dynamics discussed in \cite{qhydro} is not suitable for
applications to multiparticle production processes (among other
things because of its non covariant formulation). We would like to
fill this gap and present a fully covariant hydrodynamical model
based on $q$-statistics which can be applied to a multiparticle
production processes, especially to relativistic heavy-ion
collisions. Because of exploratory character of our paper we limit
ourselves to the $(1+1)$ dimensional case only and confront our
results with rapidity and transverse momenta distributions
obtained recently at RHIC leaving the most detailed studies of all
aspects of available experimental data for future investigations.

The hydrodynamical model of multiparticle production means, in
fact, a number of separate problems connected with the consecutive
steps of the collision process: the choice of initial conditions
summarizing the preparatory stage of collision (it should end in
some form of local thermal equilibrium), the choice of equation of
state (EoS) of the quark-gluon and/or hadronic matter being
equilibrated, further hydrodynamical evolution of this matter
assumed to form a kind of fluid, final conversion of this fluid
into observed secondaries. Because dynamical factors underlying
each step are different the resulting fluctuation patterns can
also differ, presumably  leading to the parameter $q$ changing
during the collision process. However, in the present study we
shall restrict ourself only to the case of nonextensive parameter
$q$ remaining the same in the whole collision process
\footnote{
                Notice that our analysis of multiparticle production using a $q$-hydrodynamical model
                 differs substantially from previous applications of $q$ statistics presented in
                 \cite{E1a,E1b,E2,Kq,QlQt,E3,OUWW,Lavagno,Lav-appl} because now the local thermal
                 equilibrium (in its $q$-version) is superimposed on the longitudinal
                 flow. It is therefore not clear {\it a priori} which $q$ should enter at which
                 step of the collision process. We plan to discuss this subject elsewhere.
}. 

Recently there is renewed interest in dissipative hydrodynamical
models \cite{Eckart1940,IsraelAnnPhys118,MurongaPRC69,
BaierPhysRevC73,Tsumura2007,HiscockPhysRevD31,ChaudhuriPRC74,
HeinzPRC73,KoidePRC75,DumitruPRC76}, which is prompted by the
apparent success of hydrodynamical models in describing data
obtained at RHIC \cite{HiranoPhysRevC66,TeaneyPhysRevC68,CH} and
by the recent calculations of transport coefficients of
strongly-interacting quark-gluon system using the AdS/CFT
correspondence \cite{KovtumPRL94}. The question addressed is
whether, and under which circumstances, dissipative hydrodynamics
is really needed and how it should be applied. The reason is that
formulation of the relativistic hydrodynamic equations for
dissipative fluid, suffer  from ambiguities in the form they are
written \cite{Tsumura2007}, unphysical instability of the
equilibrium state in the first order theory
\cite{HiscockPhysRevD31} and the loss of causality in the first
order equation approach \cite{KoidePRC75}, to mention a few. In
this work we argue that there is a link between dissipative
hydrodynamics ($d$-hydrodynamics) and nonextensive hydrodynamics
($q$-hydrodynamics) we are proposing, which we call {\it
nonextensive/dissipative correspondence} (NexDC). In particular,
in Section \ref{sec:meaning} we demonstrate that it is possible to
write some of the corresponding transport coefficients of the
produced matter (believed to be quark-gluon plasma (QGP)) as
(implicit) functions of the {\it nonextensivity parameter} $q$.
The merit of using the NexDC is that one can formulate and solve
the $q$-hydrodynamic equations of perfect nonextensive
hydrodynamics (or perfect $q$-hydrodynamics) in analogous way as
for the usual perfect hydrodynamics, which seems to be {\it a
priori} much easier task. Although this does not fully solve the
problems of $d$-hydrodynamics, nevertheless it allows us to extend
the usual perfect fluid approach (using only one new parameter
$q$) well behind its usual limits, namely toward the regions
reserved for dissipative approach only.

The paper is organized as follows. We start in Section
\ref{sec:formulation} with a short reminder of the nonextensive
version of kinetic theory from which nonextensive hydrodynamics is
derived in Section \ref{sec:qhydro}. Section \ref{sec:exp}
contains examples of comparisons with experimental data, whereas
in Section \ref{sec:meaning} we discuss the possible physical
meaning of the proposed $q$-hydrodynamics. We end with Section
\ref{sec:SumCon} which contains our conclusions and summary. Some
specialized topics and derivations are presented in Appendices
\ref{sec:derivation}, \ref{sec:Numerics}, \ref{sec:no-approx} and \ref{sec:shear_visco}.

\section{\label{sec:formulation}Relativistic nonextensive kinetic theory}

Following \cite{Lavagno} we start with nonextensive version of
Boltzmann equation (the metric used is: $g^{\mu\nu} =
\mbox{diag}(1,-1,-1,-1) $),
\begin{widetext}
\begin{subequations}
\begin{eqnarray}
p^{\mu} \partial_{\mu} f_q^{q}(x,p) &=& C_q(x,p) , \label{q-Boltzmann}\\
C_q(x,p)\!\!\! &=&\!\!\!  \frac{1}{2}\int \!\frac{d^3p_1}{p^0_1 }
\frac{d^3p'}{p'^0 } \frac{d^3{p_1'}}{{p_1'}^0 }
 \Big\{  h_q[f'_q,f'_{q1}]W(p',p'_1|p,p_1) -
 h_q[f_q,f_{q1}]W(p,p_1|p',p'_1)\Big\}.
\end{eqnarray}
\end{subequations}
\end{widetext}
Here $f_q(x,p)$ and $C_q(x,p)$ are $q$-versions of the,
respectively, corresponding phase space distribution function and
the $q$-collision term in which $W(p,p_1|p',p'_1)$ is the
transition rate between the two particle state with initial
four-momenta $p$ and $p_1$ and some final state with four-momenta
$p'$ and $p'_1$ whereas $h_q[f_q,f_{q1}]$ is the correlation
function related to the presence of two particles in the same
space-time position $x$ but with different four-momenta $p$ and
$p_1$, respectively. Notice two distinct features of
Eq.~(\ref{q-Boltzmann}): $(i)$ it applies to $f_q^q~(= (f_q)^q)$
rather than to $f_q$ itself and $(ii)$ in $C_q$ one assumes a new,
$q$-generalized, version of the Boltzmann molecular chaos
hypothesis \cite{Lavagno,Lima,Biro,BiroK,BiroPurcsel,SR} according
to which
\begin{eqnarray}
  h_q[f_q,f_{q1}] = \exp_q \left[ \ln_q f_q + \ln_q f_{q1}\right]  \label{q-molec-chaos}
\end{eqnarray}
where $ \exp_q(X) = [1 + (1-q)X]^{1/(1-q)}$ and $\ln_q(X)
=\left[X^{(1-q)}-1\right]/(1-q)$). Eq.~(\ref{q-molec-chaos}) is
our central point, it amounts to assuming that, instead of the
strict (local) equilibrium, {\it a kind of stationary state is
being formed, which also includes some interactions} (see
\cite{Biro,BiroK,BiroPurcsel,SR}).

With such correlation function one finds that divergence of the
entropy current, which we define as
\begin{widetext}
\begin{eqnarray}
 &&  s_q^{\mu} (x) \equiv -k_{\rm B} \int \frac{d^3p}{(2\pi\hbar)^3} \frac{p^{\mu}}{p^0}
 \Big\{  
f_q^q(x,p)\ln_q f_q(x,p) -f_q(x,p) \Big\} , \label{eq:q_entropy}
\end{eqnarray}
\end{widetext}
is always positive at any space-time point,
\begin{eqnarray}
\partial_{\mu} s_q^{\mu} (x) \ge 0, \label{eq:Htheorem}
\end{eqnarray}
(this fact is equivalent to demonstrating the validity of the
relativistic local $H$-theorem when using this $q$-entropy
current).

To get explicit form of the distribution functions $f_q(x,p)$ we
proceed now as follows \cite{Lavagno}. At first, using momentum
conservation condition in two particle collisions,
$p^{\mu}+p_1^{\mu}={p'}^{\mu}+{p'_1}^{\mu}$, we form following
collision invariant:
\begin{eqnarray}
F[\psi]=\int \! \frac{d^3p}{p^0} ~\psi(x,p) ~C_q(x,p) ~\equiv 0,
\end{eqnarray}
where $\psi(x,p)=a(x)+b_{\mu}(x)p^{\mu}$ with arbitrary functions
$a(x)$ and $b_{\mu}(x)$. We assume here that the correlation
function $h_q$ is symmetric and positive, $h_q[f,f_1]=h_q[f_1,f]
\ge 0 $, and that detailed balance holds,
$W(p,p_1|p',p'_1)=W(p',p'_1|p,p_1)$. For $a(x)\equiv 0$ and
$b_{\mu}(x)=\mbox{constant}$ one gets the $q$-version of the local
energy-momentum conservation \cite{Lavagno},
\begin{eqnarray}
      \partial_{\nu} {\cal T}_{q}^{\mu\nu}(x) =0 ,
      \label{eq:Tq-conserv}
\end{eqnarray}
with a nonextensive energy-momentum tensor defined by
\begin{eqnarray}
   {\cal T}_{q}^{\mu\nu}(x) \equiv \frac{1}{(2\pi\hbar)^3}
   \!\!\int\!\!\frac{d^3p}{p^0} ~p^{\mu}  p^{\nu} f_q^q(x,p).
 \label{eq:Tq}
\end{eqnarray}
At the same time for $a(x)=\mbox{constant}$ and $b_{\mu}(x)\equiv
0$) one gets \cite{Lavagno}
\begin{eqnarray}
\partial_{\mu}\int\!\! \frac{d^3p}{(2\pi\hbar)^3} \frac{p^{\mu}}{p^0}f_q^q(x,p) =0,
\end{eqnarray}
this implies that ($d\Omega$ stands for the corresponding phase
space volume element)
\begin{eqnarray}
\frac{d}{dt} \!\int \!\!d\Omega~ f^q(x,p) = 0 ,
\end{eqnarray}
i.e., that the normalization $Z_q \! \equiv \! \int \! d \Omega ~f_q^q(x,p) $
is conserved as well 
\footnote{
Notice that the normalization $Z_q$ and (unnormalized)
                energy density $T_q^{\mu \nu}$ are conserved independently.
                Therefore our further consideration will be concentrated only on
                getting $f_q(x,p)$. However, the $q$-dependent normalization
                is important when analyzing particle distribution
                functions because it couples the widths of distributions
                (as given by $f_q$) with their heights (as given by $Z_q$, cf. \cite{Kq}).}. 
Since the divergence of the
$q$-entropy current can also be expressed via collision invariant,
\begin{eqnarray}
 \partial_{\mu}s_q^{\mu}= \frac{1}{(2\pi\hbar)^3} ~F[\ln_q f_q(x,p)],
\end{eqnarray}
demanding that $\partial_{\mu} s_q^{\mu}(x)\equiv 0$ one finally
obtains
\begin{eqnarray}
f_q(x,p)= \Big[  1 + (1-q)\left( a(x)+b_{\mu}(x)p^{\mu} \right) \Big]^{1/(1-q)} ,
\end{eqnarray}
which represents the distribution function in a stationary state.
Setting $a(x)=0$ and $b^{\mu}(x)=-u_q^{\mu}(x)/k_{\rm B}T_q(x)$
(where $T_q(x)$ is the temperature function 
\footnote{
              One should be aware of the fact that there is still an
              ongoing discussion on the meaning of the temperature in nonextensive
              systems. However, the small values of the parameter $q$ deduced
              from data allow us to argue that, to first approximation,
              $T$ can be regarded as the hadronizing temperature in such a system.
              One must only remember that in general what we study here is not
              so much the state of equilibrium but rather some kind of stationary
              state. For a thorough discussion of the temperature of nonextensive
              systems, see \cite{Abe}.
}
) one obtains
the well known (unnormalized) Tsallis distribution function,
\begin{eqnarray}
    f_q(x,p) \!\!&=&\!\! \left[ 1-(1-q)\frac{p_{\mu}u_q^{\mu}}{k_B T_q(x)}\right]^{1/(1-q)} \nonumber \\
    \!\! &\equiv& \!\!  \exp_q\left[{-\frac{p_{\mu}u_q^{\mu}(x)}{k_B
    T_q(x)}}\right],
    \label{eq:fq}
\end{eqnarray}
where $u_q^{\mu}(x)$ should be regarded as a hydrodynamical flow
four vector (hereafter we use the convention that $\hbar=k_{\rm
B}= c=1$).

We shall now assume that the $q$-modified energy-momentum tensor
${\cal T}_q^{\mu\nu}$ can be decomposed in the usual way in terms of the
$q$-modified energy  density and pressure, $\varepsilon_q$ and $P_q$, by
using the $q$-modified flow $u_q^{\mu}$ (such that for $q\rightarrow 1$
it becomes the usual hydrodynamical flow $u^{\mu}$ and in the rest frame
of the fluid $u_q^{\mu}=(1,0,0,0)$),
\begin{subequations}
\label{eq:T_q}
\begin{eqnarray}
  {\cal T}_{q}^{\mu\nu} \!\! &=& \!\! (\varepsilon_q+P_q)u_q^{\mu}u_q^{\nu}   -P_qg^{\mu\nu} \\
  \!\!&=&\!\! \varepsilon_q u_q^{\mu}u_q^{\nu} - P_q\Delta_q^{\mu \nu}.
  \label{eq:qflow}
\end{eqnarray}
where $\Delta_q^{\mu \nu} \equiv g^{\mu \nu} - u_q^{\mu}u_q^{\nu} $.
\end{subequations}
Denoting $e\equiv p^0/T$, $z\equiv m/T$, with $g$ being degeneracy factor
depending on the type of particles composing our fluid, one gets
that in its rest frame (or in $q$-equilibrium)
\begin{widetext}
\begin{subequations}
\begin{eqnarray}
 \varepsilon_q  \!\!\!&\equiv&\!\!\! ~u_{q\mu} {\cal T}_q^{\mu\nu} u_{q\nu}  
\hspace*{3mm}
= \frac{gT_q^4}{2\pi^2}\int \!\! de \sqrt{e^2-z^2} e^2
   \left[1-(1-q)e\right]^{q/(1-q)}, \label{eq:varepsilonq}\\
 P_q \!\!\!&\equiv&\!\!\! ~-\frac{1}{3} {\cal T}_{q}^{\mu \nu} \Delta_{q\mu \nu} 
= \frac{gT_q^4}{2\pi^2} \int \!\! de \sqrt{e^2-z^2}e
   \left[1-(1-q)e\right]^{1/(1-q)},  \label{eq:Pq}\\
  s_q \!\!\! &\equiv & \!\!\! ~s_q^{\mu} u_{q\mu}  
\hspace*{11mm}
= \frac{gT_q^3}{2\pi^2} \int \!\! de \sqrt{e^2-z^2}e
  \Big\{ e[1-(1-q)e]^{q/(1-q)}  
+ [1-(1-q)e]^{1/(1-q)} \Big\}\label{eq:entropyq}
\end{eqnarray}
\end{subequations}
\end{widetext}
(notice that for $q<1$ the integration range is limited to
$z\le\epsilon\le1/(1-q)$ in order to keep the integrand positive).
It is straightforward to check that, in the baryon free case to
which we shall limit ourselves here,
\begin{subequations}
\label{eq:thermodynamical-relation}
\begin{eqnarray}
   T_q s_q = \varepsilon_q + P_q
\end{eqnarray}
and
\begin{eqnarray}
   \frac{dP_q}{dT_q}=s_q,
\end{eqnarray}
\end{subequations}
i.e., that the usual thermodynamic relations also holds for the
$q$-modified quantities.

\section{\label{sec:qhydro}The nonextensive hydrodynamical model - $q$-hydrodynamics}

\subsection{\label{sec:generalqhydro}Equations of nonextensive flow - $q$-flow}

Our starting point in formulating the $q$-hydrodynamical model is
Eq.~(\ref{eq:Tq-conserv}) with energy-momentum tensor
${\cal T}_{q}^{\mu\nu}$ given by Eq.~(\ref{eq:Tq}). Because, due to the
$q$-version of thermodynamical relations,
Eq.~(\ref{eq:thermodynamical-relation}), in our case
Eq.~(\ref{eq:Tq-conserv}) also implies conservation of
$q$-entropy,
\begin{eqnarray}
 \partial_{\mu}s_q^{\mu}=0, \label{eq:q_entropy_conservation}
\end{eqnarray}
with $ s_q^{\mu}(x) $ defined by Eq. (\ref{eq:q_entropy}), which
can also be written as
\begin{eqnarray}
 s_q^{\mu}(x) = s_q(x) u_q^{\mu}(x), \label{eq:sqcurrent}
\end{eqnarray}
we have only one general equation which, when written using
general coordinates and covariant derivatives 
\footnote{
The covariant derivatives of the vector $u^{\mu}$ and tensor
                $g^{\mu\nu}$ are defined by using the Christoffel symbol
$ \Gamma^{\nu}_{\lambda\mu}\equiv \frac{1}{2} g^{\nu\sigma}
(\partial_{\mu}g _{\sigma\lambda} +\partial_{\lambda}g_{\sigma\mu}
-\partial_{\sigma}g_{\lambda\mu})$
                 and are equal to
$ u^{\nu}_{;\mu}= \partial_{\mu} u^{\nu}
  + \Gamma^{\nu}_{\lambda\mu}u^{\lambda}\quad {\rm  and}\quad
  g^{\mu\nu}_{;\mu}= \partial_{\mu}g^{\mu\nu}
  +\Gamma^{\mu}_{\sigma\mu}g^{\sigma\nu}
  +\Gamma^{\nu}_{\sigma\mu}g^{\mu\sigma}$,   respectively.
}, 
takes
the form:
\begin{eqnarray}
  {\cal T}_{q;\mu}^{\mu\nu}=
  \left[({\varepsilon}_q+{P}_q)u^{\mu}_q u^{\nu}_q
  -{P}_qg^{\mu\nu}\right]_{;\mu}=0 .
  \label{eq:important}
\end{eqnarray}
This means that we are dealing here with {\it perfect}
$q$-hydrodynamics.

Before proceeding further some specific points of
$q$-hydrodynamics, not mentioned in the general derivation
oresented in Section \ref{sec:formulation}, must be kept in mind.
At first notice that, whereas in the usual perfect hydrodynamics
(based on BG statistics) entropy is conserved in hydrodynamical
evolution both locally and globally, in the nonextensive approach
it is conserve only locally, cf., Eq.
(\ref{eq:q_entropy_conservation}). The total entropy of the whole
{\it expanding} system is not conserved, because for two volumes
$V_{1,2}$ one finds that
\begin{eqnarray}
S_{q}^{(V_1)} + S_{q}^{(V_2)} \ne S_{q}^{(V_1\oplus V_2)},
\label{eq:S1+S2}
\end{eqnarray}
where $S_{q}^{(V)}$ are the corresponding total entropies.
Although, strictly speaking, the hydrodynamical model does not
require global entropy conservation but only its local
conservation, the above feature of $q$-hydrodynamics should be
always remembered (the consequences of this fact will be discussed
in more detail in Section \ref{sec:meaning}). Second point
concerns the causality problem. To guarantee that hydrodynamics
makes sense, there should exists some spacial scale $L$ such that
volume $L^3$ contains enough particles composing our fluid.
However, in case when there are some fluctuations and/or
correlations with some typical correlation length $l$, for which
we expect that $l> L$, one has to use nonextensive entropy
$S_q^{(L^3)}$ (cf. Eq.~(\ref{eq:S1+S2})) and its defined locally
density $s_q(x) = S_q^{(L^3)}/L^3$. When formulating the
corresponding $q$-hydrodynamics one takes, as usual,  limit $L\to
0$, in which case explicit dependence on the scale $L$ vanishes
whereas the correlation length leaves its imprint as parameter
$q$. In this sense perfect $q$-hydrodynamics can be considered as
preserving causality and nonextensivity $q$ is then related with
the correlation length $l$. One can argue that very roughly that
$q \sim l/L_{\rm eff}  \geq 1$, where $L_{\rm eff} $ is some
effective spacial scale of the $q$-hydrodynamics. Note here that
if the correlation length $l$ is compatible with the scale $L_{\rm
eff}$, i.e., $l \approx L_{\rm eff}$, one recovers condition of
the usual local thermal equilibrium and in this case the
$q$-hydrodynamics reduces to the usual (BG) hydrodynamics.

Let us now continue our presentation.  When contracted with the
velocity $u_{q\nu}$ or with the projection tensor $\Delta_{q
\lambda\nu}\equiv g_{\lambda\nu}-u_{q \lambda} u_{q\nu}$ it leads
to the following two equations:
\begin{eqnarray}
&& u_q^{\mu}\partial_{\mu} {\varepsilon}_q +
 ({\varepsilon}_q+{P}_q) u^{\mu}_{q;\mu}
  -{P}_q u_{q\nu} g^{\mu\nu}_{;\mu} =0  ,
 \label{first_equation}\\
 && ({\varepsilon}_q+{P}_q)u_q^{\mu}\Delta_{q \lambda\nu}u^{\nu}_{q;\mu} \nonumber \\
 && \hspace*{15mm} -\Delta_{q \lambda\nu} \partial^{\nu}{P}_q
 -{P}_q\Delta_{q \lambda\nu}g^{\mu\nu}_{;\mu} =0.
  \label{second_equation}
\end{eqnarray}
These are the equations to be solved now for the $(1+1)$
dimensional case. We shall assume longitudinal expansion only and
introduce proper time $\tau$ and the space-time rapidity $\eta$:
\begin{subequations}
\begin{eqnarray}
   \tau \equiv \sqrt{t^2-z^2}, \\
   \eta\equiv  \frac{1}{2}\ln \frac{t+z}{t-z}.\label{eq:taueta}
\end{eqnarray}
\end{subequations}
The corresponding metric tensor in this $(\tau$-$\eta)$ space is
$g^{\mu\nu} = {\rm diag}\left(1,-\frac{1}{\tau^2}\right)$ . The
corresponding four velocity of our fluid can be expressed by the
local fluid rapidity $\alpha_q(x)$ as
\begin{eqnarray}
u_q^{\mu}(x)=\left[ \cosh\left( \alpha_q-\eta \right),
\frac{1}{\tau}\sinh\left( \alpha_q-\eta \right) \right].
\end{eqnarray}
In this case Eq.~(\ref{first_equation}) reduces to (here $v_q\equiv
\tanh(\alpha_q-\eta)$)
\begin{eqnarray}
\frac{\partial\varepsilon_q}{\partial\tau}+\frac{v_q}{\tau}
\frac{\partial\varepsilon_q}{\partial\eta} + \left(
\varepsilon_q+P_q \right) \left\{v_q
\frac{\partial\alpha_q}{\partial\tau}+
\frac{1}{\tau}\frac{\partial\alpha_q}{\partial\eta}\right\} =0
\label{eq:3.21}
\end{eqnarray}
whereas Eq.~(\ref{second_equation}) reduces to the $q$-generalized
relativistic Euler equation (cf. Appendix \ref{sec:derivation} for
details):
\begin{eqnarray}
\left( \varepsilon_q+P_q \right)
\left\{\frac{\partial\alpha_q}{\partial\tau}+
\frac{v_q}{\tau}\frac{\partial\alpha_q}{\partial\eta}\right\}
+v_q\frac{\partial P_q}{\partial\tau}+\frac{1}{\tau}\frac{\partial
P_q}{\partial\eta}=0. \label{eq:3.22}
\end{eqnarray}
To solve these equations one needs additional input in terms of EoS,
$P_q=P_q(\varepsilon_q)$, and the choice of boundary conditions,
which we set as $v_q=0$ at $\eta=0$ (because of the symmetry
$\alpha\equiv 0$). At $\eta=0$ Eqs. (\ref{first_equation}) and
(\ref{second_equation}) reduce to
\begin{subequations}
\begin{eqnarray}
 && \frac{\partial \varepsilon_q}{\partial
    \tau}=-\frac{\varepsilon_q+P_q}{\tau} \frac{\partial \alpha}{\partial \eta}\Big|_{\eta=0} \\
 && \frac{\partial P_q}{\partial\eta}\Big|_{\eta=0} = 0.\hspace*{15mm}
\end{eqnarray}
\end{subequations}

\subsection{\label{sec:EoS}Nonextensive Equation of State - $q$-EoS}

The next important ingredient of any hydrodynamical model is an
Equation of State (EoS) defining a relation between the pressure
and the energy density, which depends on the properties of the
 \begin{figure}[h]
  \begin{center}
   \includegraphics[width=8.5cm]{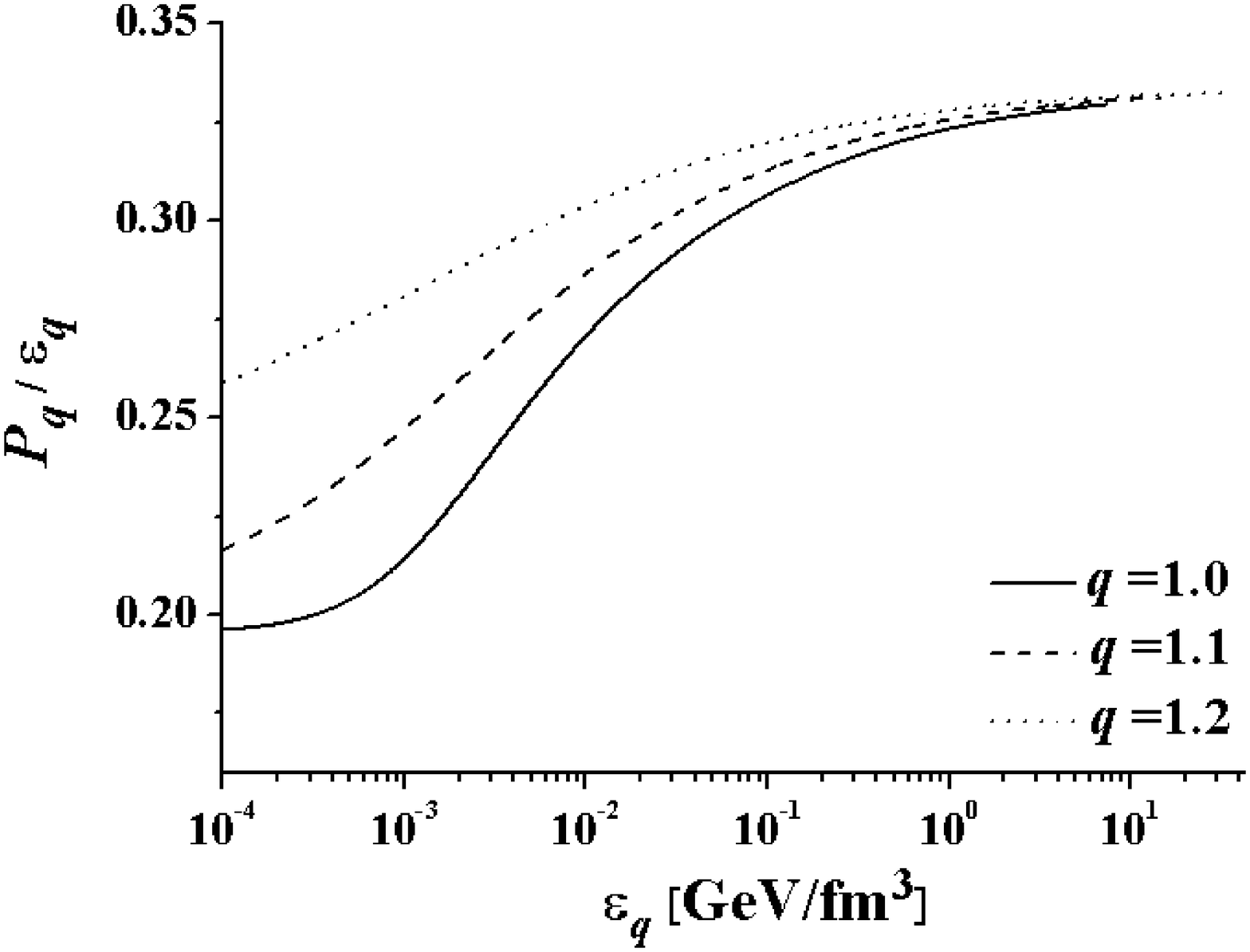}
   \caption{Illustration of EoS for the relativistic nonextensive pionic gas
            ($m = 0.14$ GeV): the ratio $P_q/\varepsilon_q$ is plotted as function
            of energy density $\varepsilon_q$ for different nonextensivity parameters
            $q$.  Notice that the $q$ dependence of EoS shows mainly at low and very
            low energy densities.}
   \label{Fig:1}
  \end{center}
\end{figure}
hadronic matter under consideration. In this work we shall only
work with EoS for the relativistic pion gas (with $m_{\pi} =0.14$
GeV) without considering different phases of hadronic matter as in
\cite{recenthydro}. The pressure $P_q$ and the energy density
$\varepsilon_q$ can be connected in the form of EoS,
$P_q(\varepsilon_q)$, using Eqs.(\ref{eq:varepsilonq}) and
(\ref{eq:Pq}). However, differently than in the usual cases of
$q=1$, the additional freedom represented by the nonextensivity
parameter $q$ makes $P_q = P_q(\varepsilon_q)$ ambiguous and one
has to additionally specify the possible variations of the
parameter $q$ during the evolution process. In what follows, we
shall assume that the parameter $q$ remains fixed during the the
whole evolution of our hadronic fluid. We therefore get different
EoS for different (but fixed) values of the parameter $q$,
examples of which are shown in Figure~\ref{Fig:1}. It displays the
ratio $P_q/\varepsilon_q$ as a function of energy density
$\varepsilon_q$ for different values of $q=1.0$, $1.1$ and $1.2$;
the temperature $T_q$ was varied in the range $0.1\div 500$ MeV. It
turns out that the $q$-dependence is confined only to the very low
energy density region (supporting therefore the previous results
on this matter in \cite{Lavagno,DragoPhysicaA344}). In the region
of interest, $\varepsilon_q \sim 0.1 \div 5.0$ GeV/fm$^3$, the
changes are very small and rapidly vanish with increasing
$\varepsilon$.

\subsection{\label{sec:incond}Nonextensive initial conditions - $q$-initial conditions}

To solve equations of $(1+1)$ $q$-hydrodynamics one has to decide
on the initial conditions from which the hydrodynamical expansion
starts. They must contain some form of the local thermal
equilibrium, which - as we assume - is established during the
collision process. According to recent estimations this can happen
very rapidly, already in the first $1$ fm of expansion, if caused
by some violent, nonperturbative mechanisms operating at this
stage 
\footnote{
                    See, for example, review by \cite{StM} and references therein.
                    We mention at this point that the so called {\it kinematic
                    thermalization} used here in which equilibration of energies is
                    due to the collisions, has been recently contrasted with the so called
                    {\it  stochastic thermalization} based on the process of erasing of
                    memory of the initial state resulting in a state of maximal entropy
                    and coinciding with the above thermal equilibrium state,
                    see \cite{CKS} and references therein.
}. 
It is thus natural to expect that there
must also exist some intrinsic fluctuations already present in
this preparatory stage of collision process which, according to
our philosophy, should be accounted for by the same
$q$-statistical approach as that used to form the
$q$-hydrodynamics. Following \cite{recenthydro,HiranoPhysRevC65}
we shall use Gaussian initial conditions interpolating between two
extreme situations, the one described by the Bjorken scaling type
model \cite{BjorkenHwa} and the other corresponding to the Landau
Model \cite{LandauSerFiz17}, but we shall modify them accordingly
by changing $\exp(X)$ to $\exp_q(X)$ (as in
Eq.~(\ref{q-molec-chaos}), it reduces to the usual Gaussian of
\cite{recenthydro} for $q=1$). As in \cite{recenthydro} initial
conditions are imposed for the energy density $\varepsilon_q$
expressed as a function of rapidity $\eta$:
\begin{eqnarray}
   && \varepsilon_{q} (\tau_0,\eta) = \varepsilon^{\rm (in)} \exp_q \!
   \left[ -\frac{ \eta^2} {2\sigma^2} 
   \right] ,
   \label{eq:inital_density}
\end{eqnarray}
In what follows we shall also require that the $q$-fluid and the
space-time rapidities coincide at $\tau_0$,
\begin{equation}
\alpha_q(\tau_0,\eta)= \eta . \label{eq:initial_vel}
\end{equation}
In all calculations presented in this paper 
we shall assume for simplicity that $\varepsilon_q$ and
$\alpha_q$ are independent of the transverse coordinate. However,
the remaining two parameters, $\varepsilon^{\rm (in)}$ and
$\sigma$ are not independent because one has to reproduce the
total energy $E_{\rm tot}$ allocated to the fluid which is fixed
by the conditions of the experiment,
\begin{eqnarray}
 E_{\rm tot}= \pi A_{\rm T}^2 \tau_0 \int \! d\eta
 ~\varepsilon_q(\tau_0,\eta),
 \label{eq:q-initial}
\end{eqnarray}
where $A_{\rm T}$ is the transverse size of the fluid, $\tau_0$ is
the initial proper time $\tau$ when the fluid starts to expand. The
$E_{\rm tot}$ can be obtained knowing the mean number of
participating nucleons $N_{\rm part}$ and the total energy loss per
participating nucleon $\Delta E$ 
\footnote{
                    Notice that, in principle, both the $N_{\rm part}$ and the $\Delta E$
                    are also fluctuating quantities but we shall not consider these fluctuations
                    here. One can argue that they are to some extent accounted for by the
                    nonextensive version of initial conditions considered here.
}
\begin{eqnarray}
E_{\rm tot}= N_{\rm part} \Delta E.
\end{eqnarray}
\begin{figure}[!htb]
\begin{center}
\includegraphics[width=8.5cm]{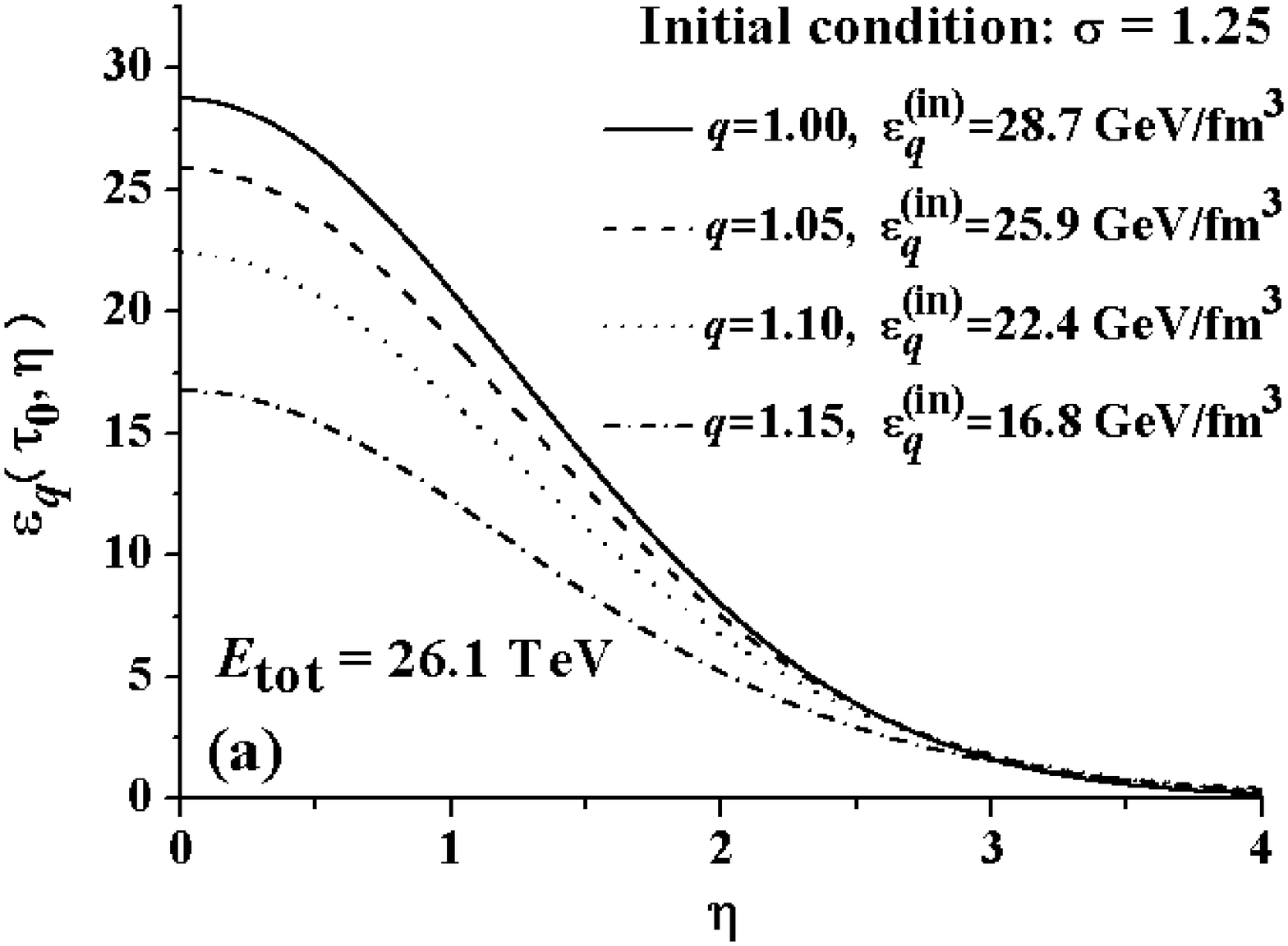} 
\includegraphics[width=8.5cm]{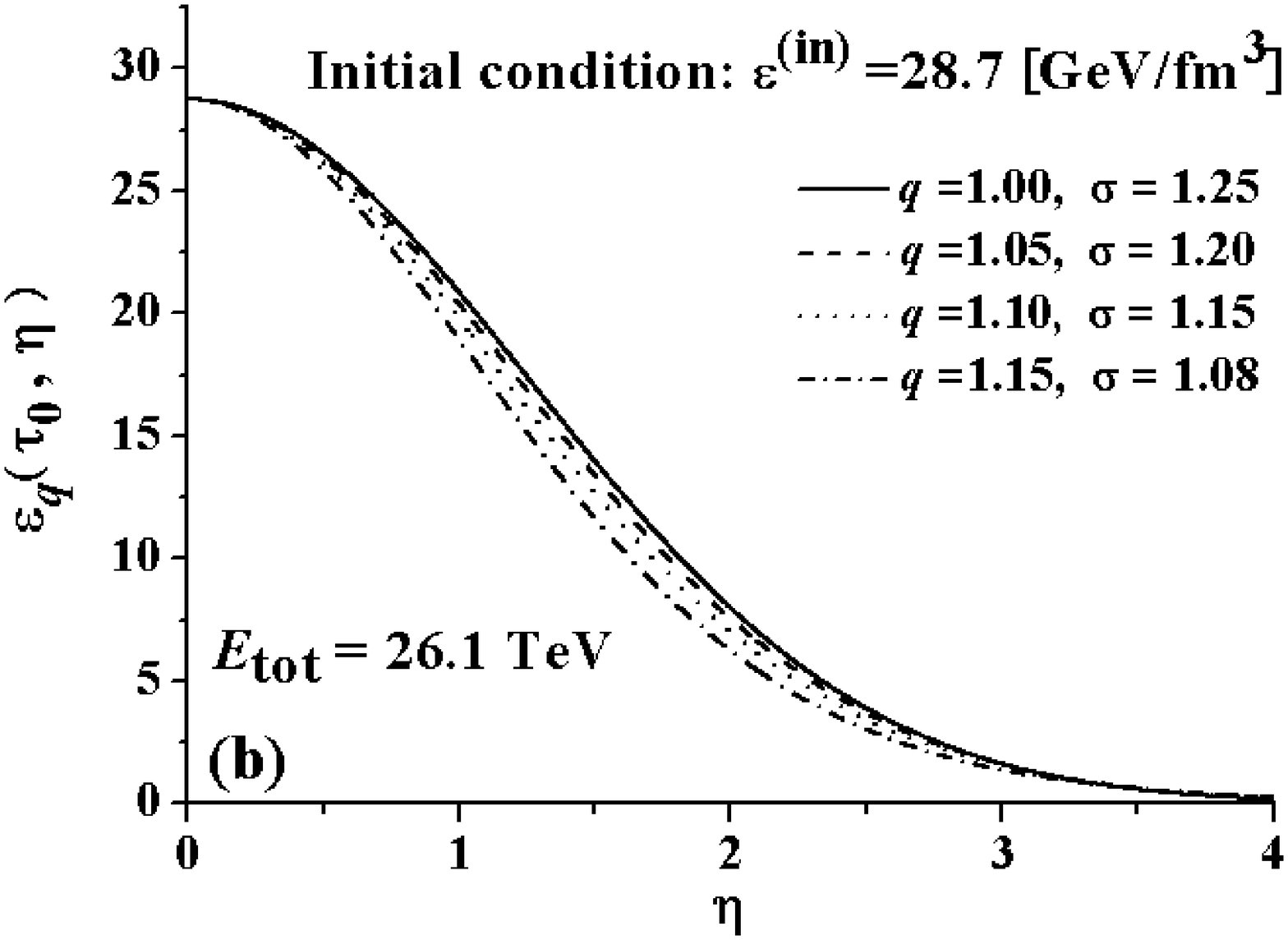}
\end{center}
\caption{Examples of two different types of initial conditions:
(a) - type $(i)$ initial conditions with fixed $\sigma = 1.25$
and $\varepsilon^{\rm (in)}$ varying according to
Eq.~(\ref{eq:q-initial}); (b) - type $(ii)$ initial conditions
with fixed $\varepsilon^{\rm (in)} = 28.7$ [GeV/fm$^{3}$] (this is the
value corresponding to the $q=1$ situation in (a)) and $\sigma$
varying according to Eq.~(\ref{eq:q-initial}).}
\label{Fig:2}
\end{figure}
The possible initial conditions vary therefore between two
extremal situations (cf., Fig.~\ref{Fig:2}):
\begin{itemize}
\item[$(i)$] The width $\sigma$ in Eq.~(\ref{eq:inital_density})
is assumed to be fixed and kept constant, but its distribution to
vary by changing $\varepsilon^{\rm (in)}$ to reproduce the fixed total
energy $E_{\rm tot}$ when the nonextensive parameter $q$ changes.
\item[$(ii)$] The maximum energy density $\varepsilon^{\rm (in)}$ in
Eq~.(\ref{eq:inital_density}) is assumed to be fixed and kept
constant, but its distribution to vary by changing $\sigma$ to
reproduce the fixed total energy $E_{\rm tot}$ when the
nonextensive parameter $q$ changes.
\end{itemize}
\begin{table*}[hbt!]
\caption{List of parameters of the initial conditions used in
Fig.~\ref{Fig:2}. The initial temperature $T_{\rm in}\equiv
T(\tau_0,\eta$=0$)$ is shown for two types of EoS: for the
relativistic nonextensive pion gas for some selected values of $q
\ge 1$ and for the usual BG pion gas with $q=1$.}
\newcommand{\lw}[1]{\smash{\lower2.0ex\hbox{#1}}}
\newcommand{\lww}[1]{\smash{\lower0.8ex\hbox{#1}}}
\begin{center}
\begin{tabular}{|c|c|c|c|c|c|c|c|c|c|c|} \hline
\multicolumn{11}{|l|}{Initial condition: $\sigma$=1.25 fixed}
\\ \hline $E_{\rm tot}$ &$\varepsilon^{\rm (in)}$&\multicolumn{4}{c|} {$\sigma$} &\lw{EoS}
&\multicolumn{4}{c|}{$T_{\rm in}$ [GeV]}
\\ \cline{3-6}\cline{8-11}
[TeV]&[GeV/fm$^{3}$]  & $q$=1.00 & $q$=1.05 & $q$=1.10 & $q$=1.15
& &$q$=1.00&$q$=1.05 &$q$=1.10 &$q$=1.15 \\ \hline
\lw{26.1}&\lw{27.8} & & 1.20&1.15 & 1.08 & nonex.$\pi$ gas
&&0.648&0.591&0.531 \\  \cline{4-7}\cline{9-11}
             &             &\multicolumn{4}{c|}{1.25}  & BG $\pi$ gas &\multicolumn{4}{c|}{0.702} \\
\hline
\end{tabular}
\\
\begin{tabular}{|c|c|c|c|c|c|c|c|c|c|c|} \hline
\multicolumn{11}{|l|}{Initial condition:
$\varepsilon^{\rm (in)}$=28.7 GeV/fm$^{3}$ fixed}
\\ \hline $E_{\rm tot}$ &\hspace*{5.5mm} $\sigma$ \hspace*{5.5mm} &\multicolumn{4}{c|} {$\varepsilon^{\rm (in)}$
[GeV/fm$^3$ ] } &\lw{EoS} &\multicolumn{4}{c|}{$T_{\rm in}$ [GeV]}
\\ \cline{3-6}\cline{8-11}
[TeV]&  & $q$=1.00 & $q$=1.05 & $q$=1.10 & $q$=1.15  &
&$q$=1.00&$q$=1.05 &$q$=1.10 &$q$=1.15 \\ \hline
\lw{26.1}&\lw{1.25} & & 25.9 &22.4 & 16.8 & nonex.$\pi$ gas
&&0.631&0.556&0.464 \\  \cline{4-7}\cline{9-11}
             &             &\multicolumn{4}{c|}{27.8}  & BG $\pi$ gas &\multicolumn{4}{c|}{0.702} \\ \hline
\end{tabular}
\end{center}
\label{tab.IIIC1}
\end{table*}
\begin{figure*}[hbt!]
  \begin{center}
  \includegraphics[width=8.5cm]{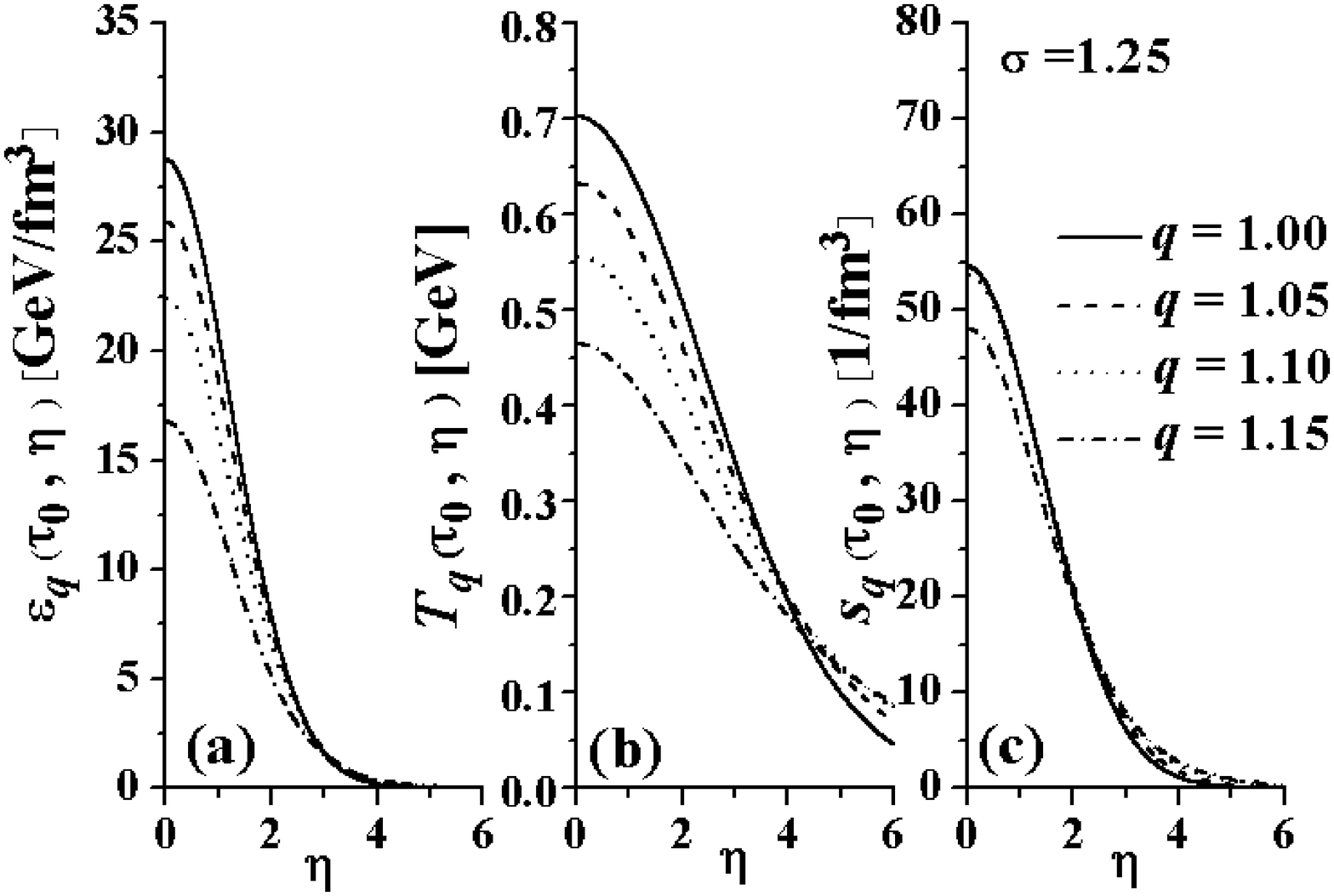}
  \includegraphics[width=8.5cm]{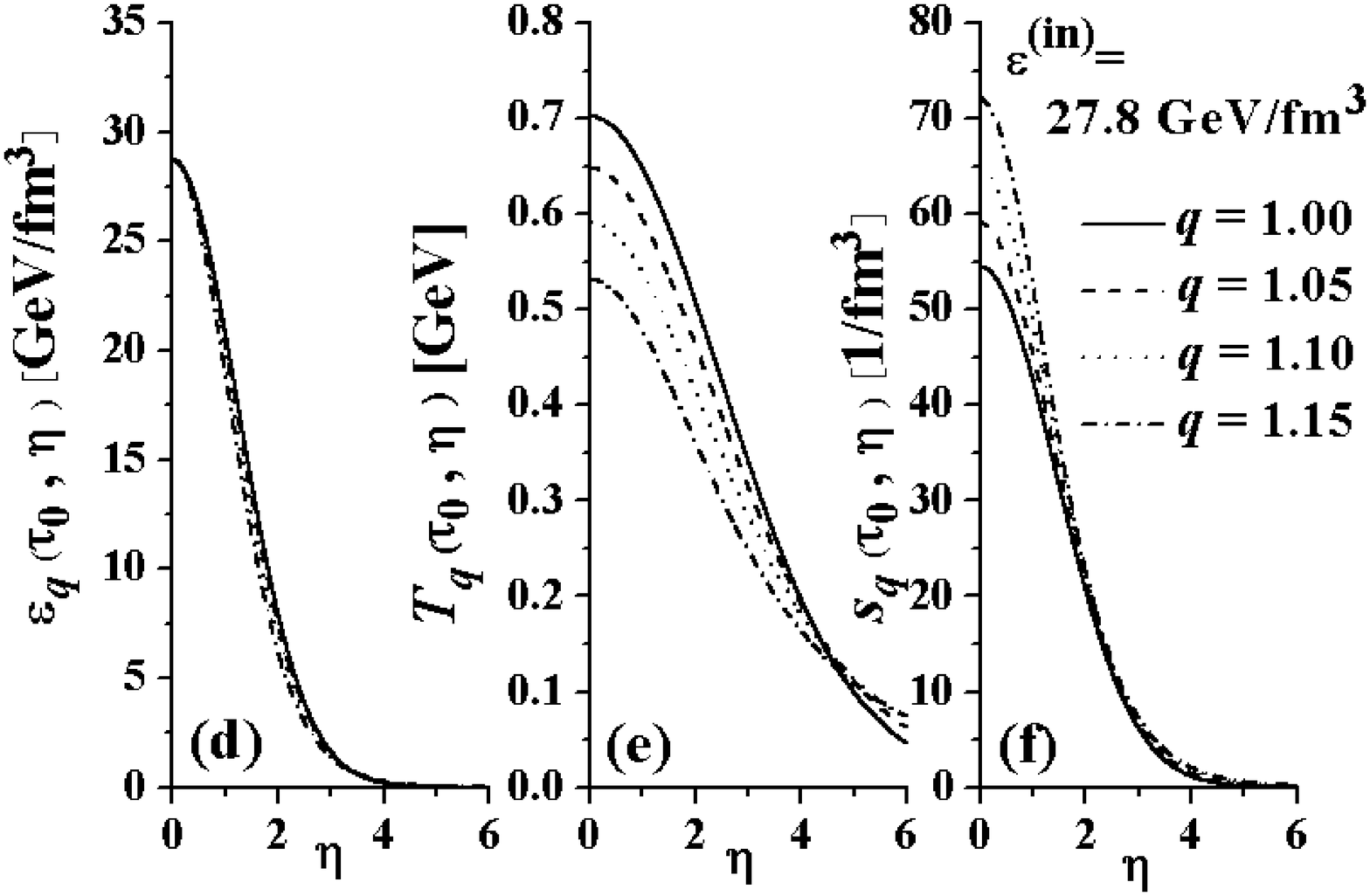}
   \caption{Dependence of the initial conditions on the parameter $q$.
            The initial conditions for the energy density $\varepsilon_q(\tau_0,\eta)$ (panels
            (a) and (d)), the corresponding temperature $T_q(\tau_0,\eta)$ (panels (b) and (e))
            and the entropy density $s_q(\tau_0,\eta)$ (panels (c) and (f))
            are plotted for different values of $q$ and for two types of initial conditions:
            $(i)$ when the initial width $\sigma=1.25$ remains fixed
            (panels (a)-(c)) and $(ii)$ when the initial maximal energy density
            $\varepsilon^{\rm (in)} = 27.8$ GeV/fm$^{3}$ remains fixed (panels (d)-(f)).
            In both cases we assume that $E_{\rm tot}$ is the same for all $q$'s and
            equal to $E_{\rm tot} = 26.1$ TeV whereas $\tau_0$ in Eq~.(\ref{eq:inital_density})
            is put equal to $\tau_0 = 1.0$ fm. Temperatures $T_q$ for different
            values of $q$ displayed on panels (b) and (e)
            are determined by solving Eq.~(\ref{eq:varepsilonq})
            whereas the entropy densities  $s_q$ displayed on panels (c) and (f) are obtained
            from Eq.~(\ref{eq:entropyq}) using values of $T_q$ displayed in panels (b) and (e).}
   \label{Fig:3}
  \end{center}
\end{figure*}
\begin{table*}[htb!]
\caption{The values of energy density $\varepsilon_q$, entropy
density $s_q$ and temperature $T_q$ at $\eta =0.0$ and $\eta = 3.0$
and at $\tau = 5.0$ and $25.0$ fm for $q =1.0$ and $q = 1.1$ for
two extremal types of initial conditions: with fixed $\sigma =
1.25$ (upper panel) and with fixed
$\varepsilon^{in}$=27.8GeV/fm$^{3}$ (lower panel). }
\label{tab.IIID2}
\newcommand{\lw}[1]{\smash{\lower2.0ex\hbox{#1}}}
\newcommand{\lww}[1]{\smash{\lower0.8ex\hbox{#1}}}
{\footnotesize
\begin{center}
\begin{tabular}{|c|c||c|c|c|c|c|c||c|c|c|c|c|c||c|c|c|c|c|c|}\hline
\multicolumn{20}{|l|}{Initial condition: $\sigma$=1.25 fixed} \\ \hline
\multicolumn{2}{|c||}{ }
&\multicolumn{6}{c||}{$\varepsilon_q$~[GeV/fm$^3$]}
&\multicolumn{6}{c||}{$T_q$~[GeV]} &\multicolumn{6}{c|}{$s_q$
~[1/fm$^{3}$]}
\\ \hline \hline
\multicolumn{2}{|c||}{$\tau$} &\multicolumn{2}{c|}{1 fm}
&\multicolumn{2}{c|}{5 fm} & \multicolumn{2}{c||}{25 fm}&
  \multicolumn{2}{c|}{1 fm} &\multicolumn{2}{c|}{5 fm} & \multicolumn{2}{c||}{25 fm}&
  \multicolumn{2}{c|}{1 fm} &\multicolumn{2}{c|}{5 fm} & \multicolumn{2}{c|}{25 fm}
\\ \hline
\multicolumn{2}{|c||}{$\eta$}& 0.0&3.0&0.0&3.0&0.0&3.0
                                       & 0.0&3.0&0.0&3.0&0.0&3.0
                                       &0.0&3.0&0.0&3.0&0.0&3.0 \\ \hline
\lw{$q$} &1.00&
  28.7   & 1.61 &
  2.80   & 0.293&
  0.249 & 0.060&
  0.702 & 0.343&
  0.393 & 0.225 &
  0.216 & 0.153&
  54.5   & 6.24 &
  9.46   & 1.72&
  1.52   & 0.513
 \\ \cline{2-20}
           & 1.10&
  22.4   & 1.78 &
  2.19   & 0.285&
  0.196 & 0.053&
  0.556 & 0.296 &
  0.311 & 0.189 &
  0.171 & 0.124 &
  53.7 & 8.00 &
  9.34 & 2.01&
  1.51 & 0.553
\\
\end{tabular}
\\

\noindent
\begin{tabular}{|c|c||c|c|c|c|c|c||c|c|c|c|c|c||c|c|c|c|c|c|}\hline
\hline \multicolumn{20}{|l|}{Initial condition:
$\varepsilon^{\rm (in)}$=28.7 GeV/fm$^{3}$ fixed} \\ \hline
\multicolumn{2}{|c||}{ }
&\multicolumn{6}{c||}{$\varepsilon_q$~[GeV/fm$^3$]}
&\multicolumn{6}{c||}{$T_q$~[GeV]} &\multicolumn{6}{c|}{$s_q$
~[1/fm$^{3}$]}
\\ \hline \hline
\multicolumn{2}{|c||}{$\tau$} &\multicolumn{2}{c|}{1 fm}
&\multicolumn{2}{c|}{5 fm} & \multicolumn{2}{c||}{25 fm}&
  \multicolumn{2}{c|}{1 fm} &\multicolumn{2}{c|}{5 fm} & \multicolumn{2}{c||}{25 fm}&
  \multicolumn{2}{c|}{1 fm} &\multicolumn{2}{c|}{5 fm} & \multicolumn{2}{c|}{25 fm}
\\ \hline
\multicolumn{2}{|c||}{$\eta$}& 0.0&3.0&0.0&3.0&0.0&3.0
                                       & 0.0&3.0&0.0&3.0&0.0&3.0
                                       &0.0&3.0&0.0&3.0&0.0&3.0 \\ \hline
\lw{$q$} &1.00&
  28.7  & 1.61&
  2.80  & 0.293&
  0.249& 0.060&
  0.702&0.343&
  0.393&0.225&
  0.216&0.153&
  54.5 & 6.24&
  9.46 & 1.72&
  1.52 & 0.513
 \\ \cline{2-20}
           & 1.10&
  28.7  & 1.53&
  2.72  & 0.272&
  0.235 & 0.057&
  0.591 & 0.285&
  0.329 & 0.186&
  0.179 & 0.127&
  64.7 & 7.14&
  11.0 & 1.94&
  1.73& 0.587
\\ \hline
\end{tabular}
\end{center}
}
\end{table*}
We would like to stress at this point that such $q$-dependent
initial conditions introduce a completely new element to
hydrodynamical models, not discussed previously. The real
situation will interpolate in an {\it a priori} unknown manner
between these two extremes, therefore, in what follows, we shall
restrict ourself only to them. As one can see in
Fig.~\ref{Fig:2}, whereas the first extreme introduces
sizable $q$-dependence, the second one leads to only minor
effects. In both cases, increasing the value of $q$ results, as
expected \cite{Kq}, in the enhancement of tails for large values
of $\eta$. Following \cite{recenthydro} our calculations were
performed for Au+Au collisions at $\sqrt{s_{\rm NN}} = 200$ GeV,
using results reported by the BRAHMS experiment
\cite{BRAHMSPhysRevLett93}, with $E_{\rm tot} =26.1$ TeV , $N_{\rm
part} = 357$, $\Delta E=73\pm 6$ GeV and with
$A_{\rm T} = 6.5$ fm, $\tau_0 = 1.0$ fm, see
Table.\ref{tab.IIIC1}. In Fig.~\ref{Fig:3} are shown, the initial
conditions for the, respectively, energy density
$\varepsilon_q(\tau_0,\eta)$, entropy density $s_q(\tau_0,\eta)$
and temperature $T_q(\tau_0,\eta)$ in the case of $\sigma = 1.25$
fixed (panels (a)-(c)) and $\varepsilon^{\rm (in)} = 27.8$ GeV/fm$^{3}$
fixed (panels (d)-(f)) and reproducing the initial energy $E_{\rm
tot} = 26.1$ TeV. We start with
$\varepsilon_q$ (panels (a) and (d) of Fig.~\ref{Fig:3}) which is given
by Eq.~(\ref{eq:inital_density}), use it to solve Eq.~(\ref{eq:varepsilonq})
and find $T_q(\tau_0)$ (panels (b) and (e) of Fig.~\ref{Fig:3}) and eventually obtain
$s_q(\tau_0,\eta)$ using these results and Eq.~(\ref{eq:entropyq})
(panels (c) and (f) of Fig.~\ref{Fig:3}).

\begin{figure*}[htb!]
  \begin{center}
 \includegraphics[width=8.5cm]{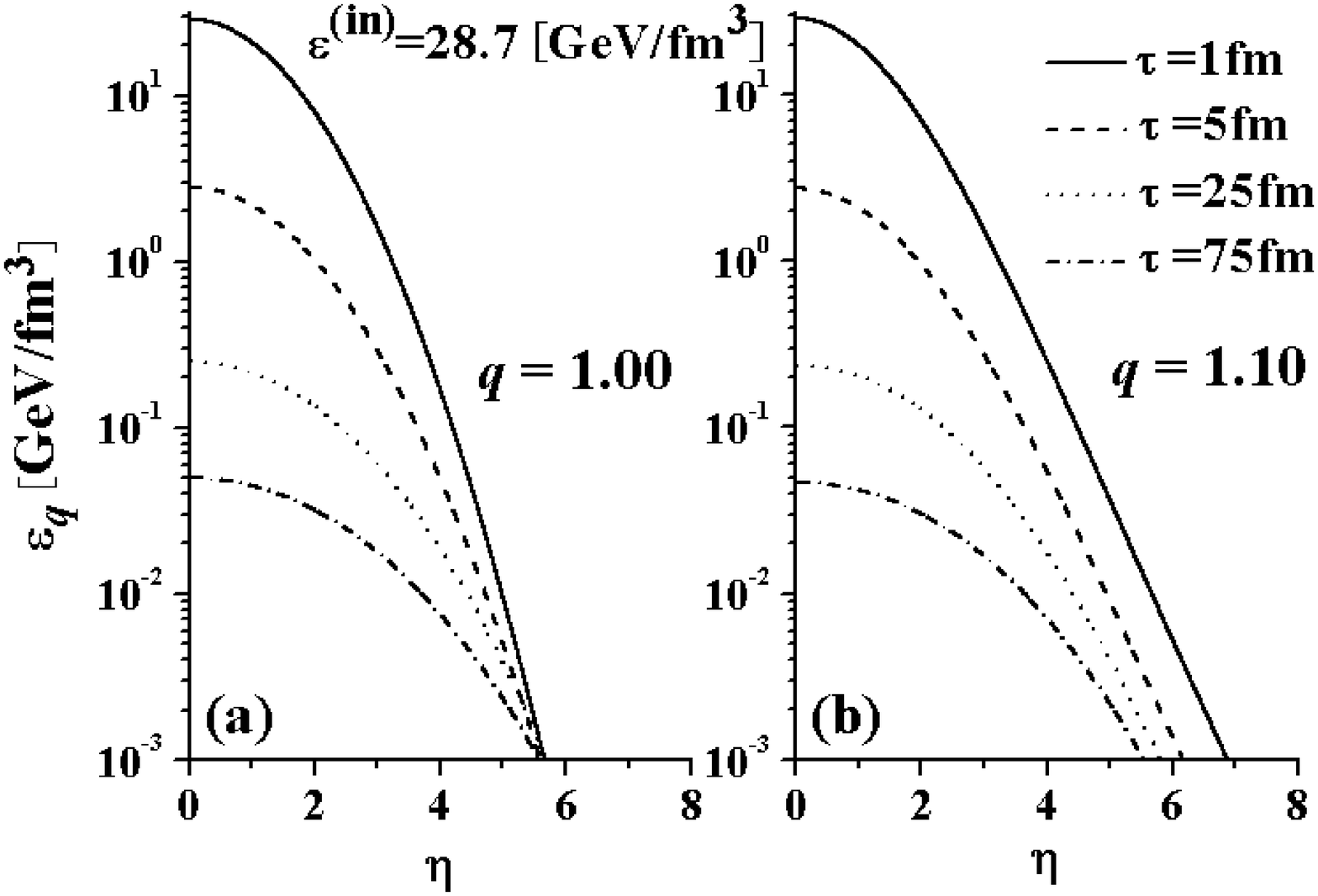}
 \includegraphics[width=8.5cm]{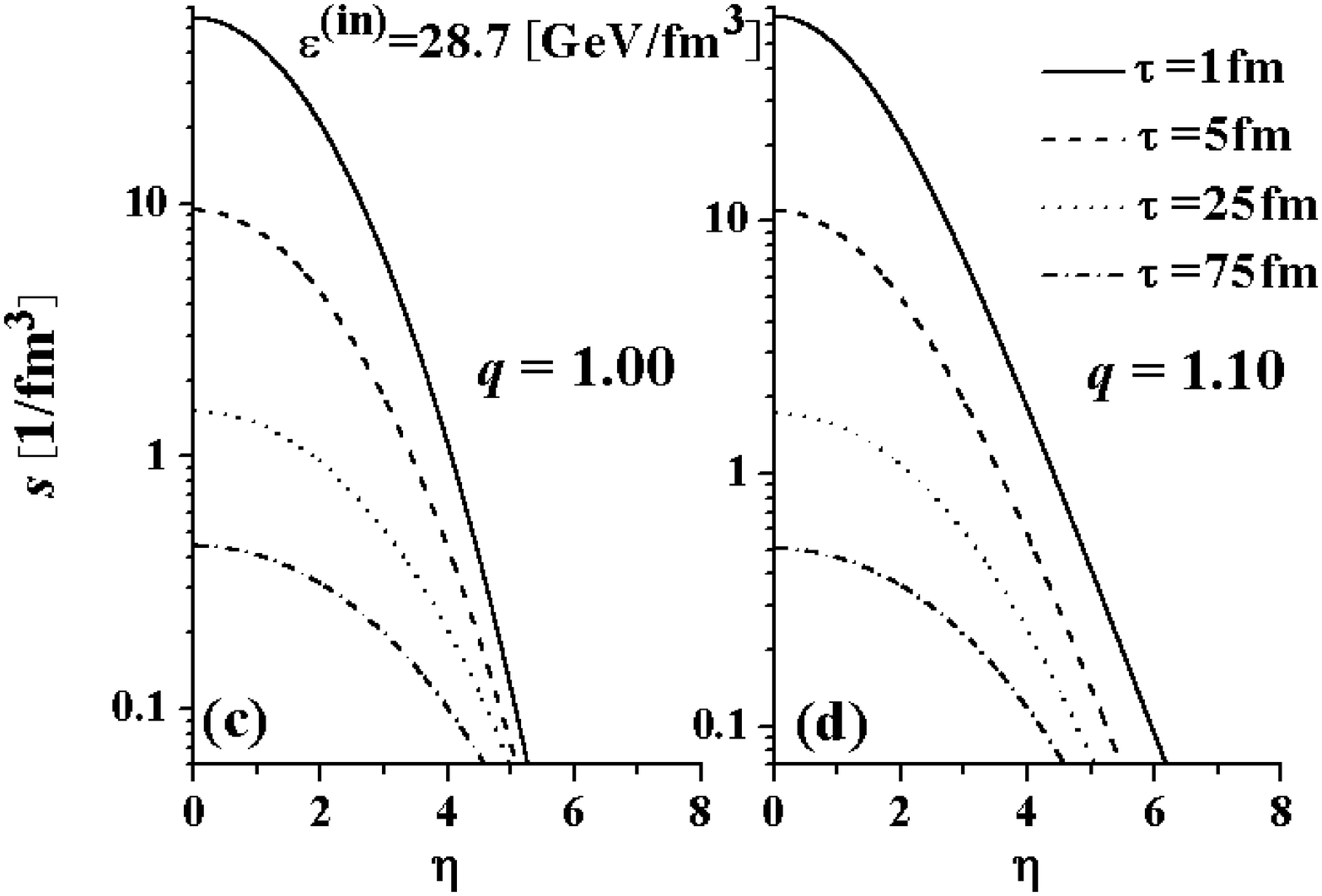}
 \includegraphics[width=8.5cm]{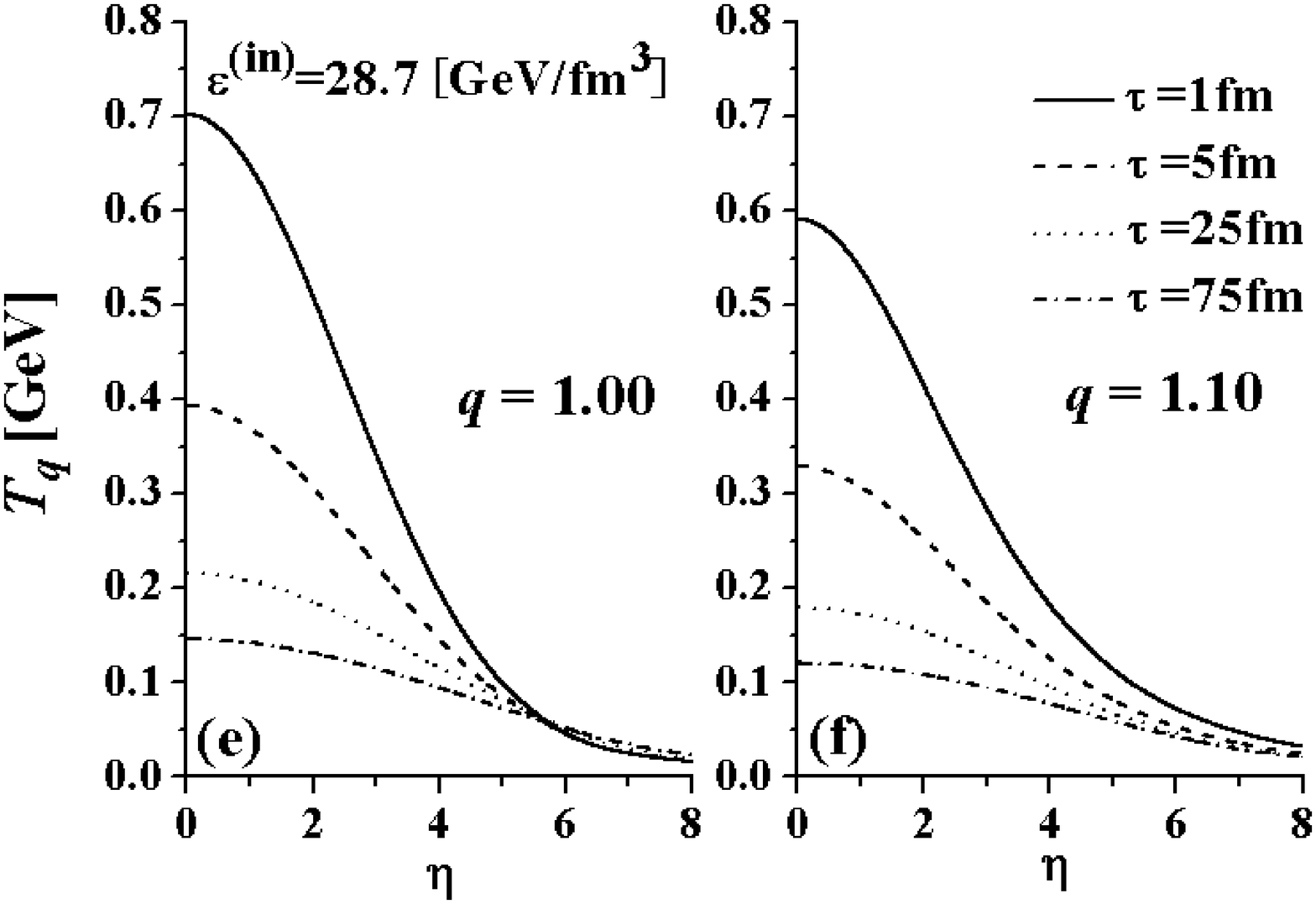}
  \caption{Profiles of the energy density $\varepsilon_q$ (panels (a) and
(b)), entropy density $s_q$ (panels (c) and (d)) and
temperature $T_q$ (panels (e) and (f)) as functions of $\eta$
calculated for different proper times $\tau$ and different
nonextensivity parameters $q$ by using initial conditions with
fixed $\varepsilon^{\rm (in)}$=28.7~GeV/fm$^{3}$. We also obtain
similar results for $\varepsilon_q, T_q$ and $s_q$ with
$\sigma=1.25$ fixed initial condition case.}\label{Fig:4}
  \end{center}
\end{figure*}

\subsection{\label{sec:Examples_of_evolution}Examples of $q$-hydrodynamical evolution
            of different thermodynamical quantities}

Let us now demonstrate examples of $q$-hydrodynamical evolution of
different thermodynamical quantities and of the fluid rapidity.
Calculations were performed using the method presented in Appendix
\ref{sec:Numerics}. In Fig~\ref{Fig:4} we
present the evolution of the, respectively, energy density
$\varepsilon_q$, temperature $T_q$ and entropy density $s_q$ using
initial conditions discussed in Section \ref{sec:incond} (the
exact values of relevant parameters for both types of initial
conditions are listed in Table \ref{tab.IIID2}). One can see that
the initial functional forms of $\varepsilon_q(\tau,\eta)$,
$T_q(\tau,\eta)$ and $s_q(\tau,\eta)$ generally follow their
original Gaussian shapes assumed at the initial time $\tau_0$ for
$q =1.0$, $1.05$ and $1.1$. On the other hand, during the whole
hydrodynamical evolution both the energy density $\varepsilon_q$
and the temperature $T_q$ calculated for $q > 1$ are smaller than
those for $q=1$ for $\tau > \tau_0 =1$ fm (see Table
\ref{tab.IIID2}). That is even true for the initial condition with
fixed $\sigma = 1.25$, for which the initial energy density
$\varepsilon_{q=1.1}(\tau_0,\eta=3) >
\varepsilon_{q=1.0}(\tau_0,\eta=3)$, in which case, after the
$q$-hydrodynamical evolution is completed, one observes that
$\varepsilon_{q=1}  > \varepsilon_{q>1}$. The same trend is also
observed for the temperature, i.e., $T_{q=1}  > T_{q>1}$ for all
$\tau$ and $\eta$. However, the corresponding entropy density
$s_q$ evolves differently:
for both type of initial conditions and any $\eta$,
inequality relations between $s_{q=1.1}(\tau,\eta)$ and $s_{q=1.0}(\tau,\eta)$
given at initial $\tau=\tau_0$ are preserved during hydrodynamical evolution. 

\begin{figure}[hbt!]
  \begin{center}
 \includegraphics[width=9.0cm]{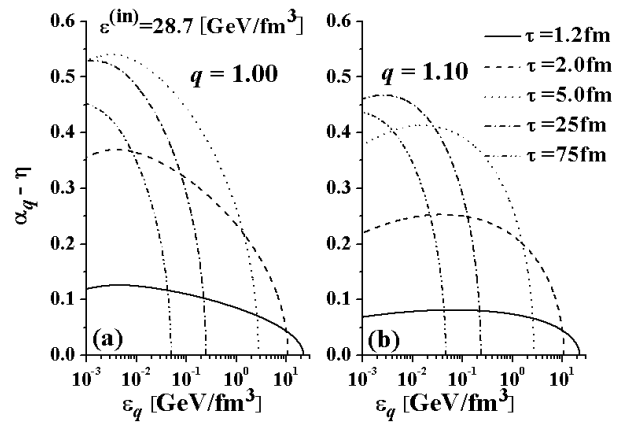}
  \caption{(a) - evolution of the fluid rapidity $\alpha_q$ (presented as $\alpha_q-\eta$)
  as a function of the energy density $\varepsilon_q$  for different values of  $\tau$ and for
  different nonextensivity parameters $q$ with fixed $\varepsilon^{\rm (in)}$=
 28.7 GeV/fm$^{3}$. (b) - similar results for $\alpha_q-\eta$
 using $\sigma=1.25$ fixed initial condition case.}
 \label{Fig:5}
 \end{center}
 \end{figure}

In what concerns the fluid rapidity $\alpha_q$, it is always set to be
equal to $\alpha_q(\tau_0,\eta) \equiv \eta $ at $\tau=\tau_0$.
However, the pressure gradient, which is characteristic to
Gaussian type initial conditions applied here, accelerates the
fluid, therefore $\alpha_q$ evolves with time $\tau$ (actually,
this is true even for $q$=1), see Fig.~\ref{Fig:5}. In these
figures the fluid rapidity $\alpha_q$ (actually its deviation from
the rapidity $\eta$, $\alpha_q-\eta$) is shown as a function of
$\tau$ and the corresponding energy density $\varepsilon_q$.
Notice that $\alpha_q-\eta\equiv 0$ at $\tau_0$ for the whole
$\eta$ space (i.e. for all region of the $\varepsilon_q$). As
shown in Fig.~\ref{Fig:5}, the fluid rapidity $\alpha_q$ grows
during the hydrodynamical expansion from its initial value
$\alpha(\tau_0,\eta) = \eta$. One can observe that $\alpha_q$ for
$q>1$ is decelerated compared to the usual hydrodynamic expansion
(i.e., $\alpha_{q=1} > \alpha_{q>1}$).

To summarize this part: one observes that nonextensive fluid
($q$-fluid with $q>1$) evolves more slowly than the ideal fluid
(with $q=1$).

\subsection{\label{sec:freeze}Freezeout surface and single particle spectra}

\begin{figure}[hbt!]
\begin{center}
\includegraphics[width=8.5cm]{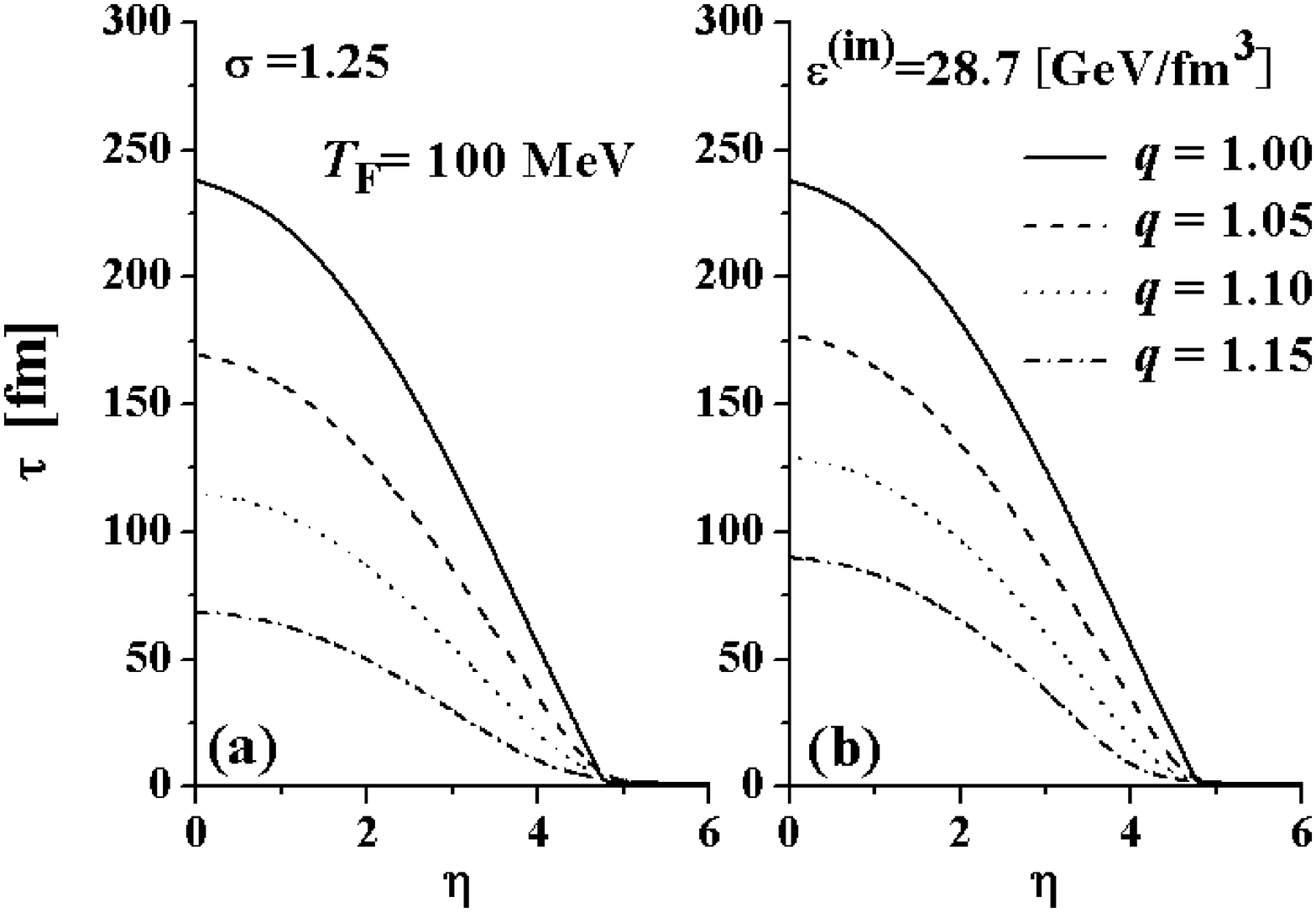}
\caption{ Examples of freezeout surfaces of constant temperature
$T_{\rm F}$=100~MeV calculated for different values of parameter
$q$ and for different initial conditions: with constant
$\sigma=1.25$ (a) and with constant energy density
$\varepsilon_{\rm F}$=9.82$\times 10^{-3}$~GeV/fm$^{3}$~(b).
}\label{Fig:6}
\end{center}
\end{figure}

We now present examples of single particle spectra emerging from
our approach. We shall follow the simplest possibility in which
they are expressed as an integral of the phase space particle
density over a freeze-out surface $\Sigma_{\rm f}$
\cite{CooperPhysRevD10},
\begin{eqnarray}
   E\frac{d^3N}{dp^3} \!\!&=&\!\! \frac{d^3N}{m_{\rm T}dm_{\rm T} dy d\phi} \nonumber \\
   \!\!&=&\!\!
   \frac{g}{(2\pi)^3} \int_{\Sigma_{\rm f}} \!\!\!
   d\sigma_{\mu}(x)~ p^{\mu} f_{\rm eq}(x,p) .
   \label{EurPhysJC.eq.6}
\end{eqnarray}
In the $\tau$-$\eta$ metric the surface element of $\Sigma_{\rm
f}$ given by
\begin{eqnarray}
   d\sigma_{\mu}= \left( d\sigma_{\tau}, d\sigma_{\eta} \right)=
   A_{\rm T} ~\tau d\eta\left( 1,-\frac{n_{\eta}}{n_{\tau}}
   \right),
\end{eqnarray}
where $n_{\mu}$ is the normal covariant vector of the isothermals,
\begin{eqnarray}
  n_{\mu} = \left( n_{\tau},n_{\eta} \right) =
  \left( -\frac{\partial T}{\partial \tau}, -\frac{\partial T}{\partial\eta}
  \right),
  \label{normal_vector}
\end{eqnarray}
and  $A_{\rm T}$ is the transverse area of the generated fluid. In
all examples of applications to Au+Au collisions discussed in this
paper we use $A_{\rm T} = 6.5$ fm. The momentum of the produced
particle in the $\tau$-$\eta$ metric is given  by
\begin{eqnarray}
   p^{\mu} = \left[ m_{\rm T}\cosh(y-\eta), \frac{1}{\tau}m_{\rm T}\sinh(y-\eta) \right]
\end{eqnarray}
where $y$ is the observed rapidity (after freezeout). Using these
expressions, the single particle density is given by:
\begin{widetext}
\begin{eqnarray}
   E\frac{d^3N}{dp^3}  = \frac{d^3N}{m_{\rm T}dm_{\rm T}dy} =
     \frac{g A_{\rm T}^2 }{4\pi} \!\int \!\!d\eta  ~\tau_{\rm f}(\eta) ~\left[
  m_{\rm T}  \cosh(y-\eta) -\frac{1}{\tau} \frac{n_{\eta}(\eta)}{n_{\tau}(\eta)} m_{\rm T}\sinh(y-\eta)
  \right]~f_q(y,\eta),
\label{single_particle_density}
\end{eqnarray}
\end{widetext}
where
\begin{eqnarray}
  f_q(y,\eta)=
  \left[1-(1-q)\frac{m_{\rm T} \cosh(y-\eta)  }{T_{\rm F}}~\right]^{\frac{1}{1-q}}\qquad
  \label{eq:fq_CF}
\end{eqnarray}
and $T_{\rm F}$ is the freezeout temperature which is given by the
corresponding freezeout energy density $\varepsilon_{\rm F}$. In
principle, the freezeout surface can be defined either as the
surface of constant temperature $T_{\rm F}$, or as the surface of
constant energy density $\varepsilon_{\rm F}$, or, finally, as the
surface of constant entropy density $s_{\rm F}$ (cf. Table
\ref{tab.IIIE3}). They all coincide in the usual extensive case
($q$=1). In Fig. \ref{Fig:6} we show as example freezeout surfaces
(calculated for different values of parameter $q$ and for
different initial conditions) for $T_{\rm F} = 100$ MeV. One
observes a quite strong $q$ dependence of the freezeout surface
characteristics on these parameters. These dependencies are much
weaker when calculated for surface of constant energy density and
even weaker for constant entropy density (not shown here
explicitly). Note that values of $T_{\rm F}$ corresponding to
freezeout conditions set by fixing $\varepsilon_{\rm F}$ or
$s_{\rm F}$ now depend on the parameter $q$ (see Table
\ref{tab.IIIE3}).

\begin{figure}[hbt!]
\begin{center}
\includegraphics [width=8.5cm]{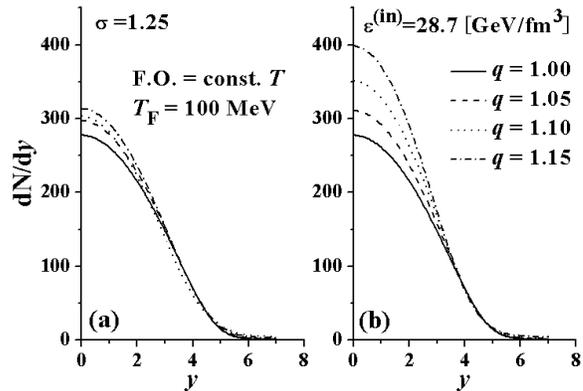}
\includegraphics [width=8.5cm]{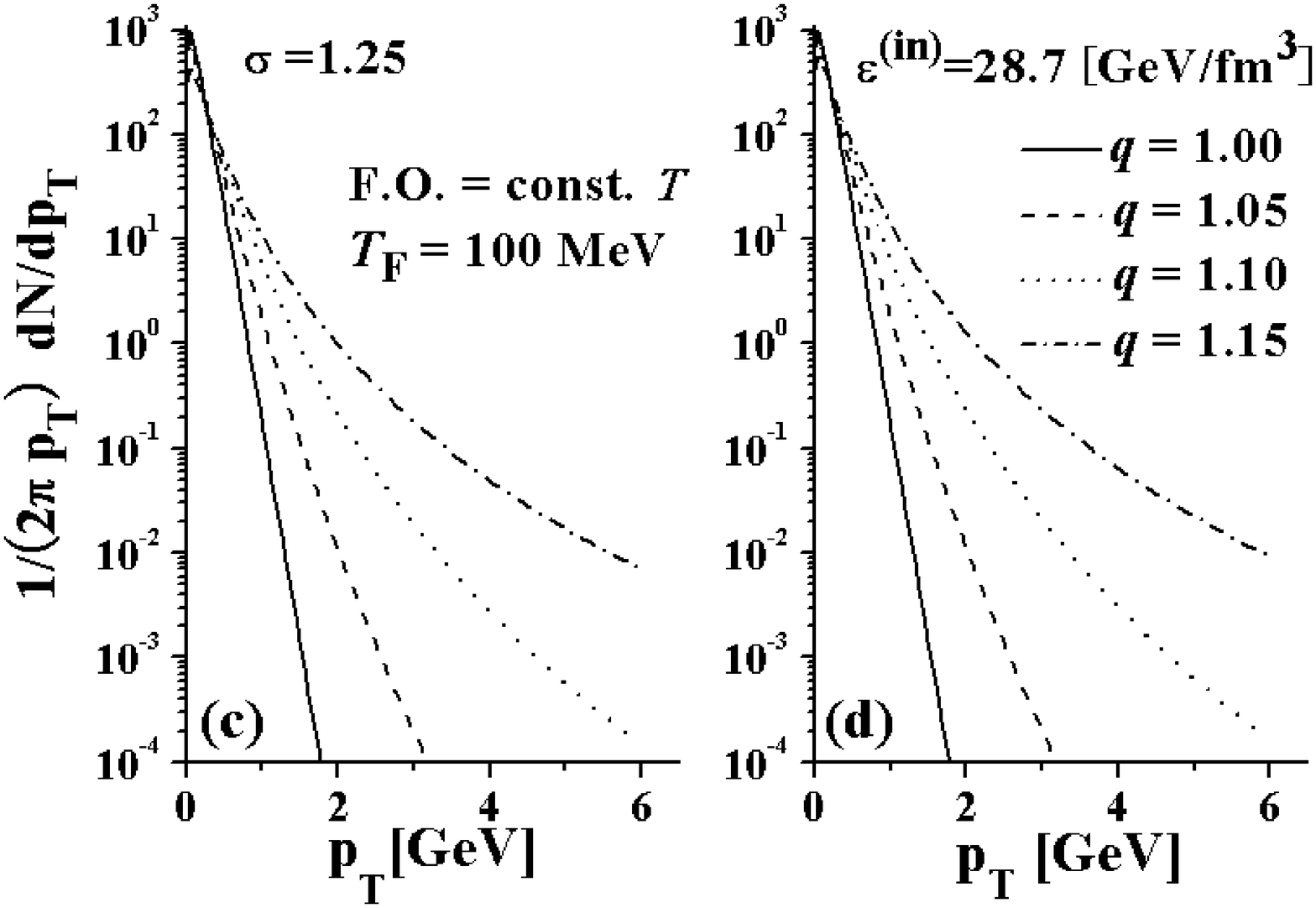}
\caption{ $dN/dy$ and $p_{\rm T}$ spectra (panels (a)-(b) and
(c)-(d), respectively) obtained from the $q$-hydrodynamical
evolution with constant $T_{\rm F} =$100 MeV and for different
values of parameter $q$ and for both types of initial conditions:
with fixed $\sigma = 1.25$ (panels (a) and (c)) and for fixed
$\varepsilon^{\rm (in)} = 27.8$ GeV/fm$^{3}$ (panels (b) and
(d) ). Rapidity spectra are obtained by integrating
Eq.(\ref{single_particle_density}) over $p_{\rm T}\in (0,6.0)$
GeV/c whereas $p_{\rm T}$ spectra are obtained by integrating
Eq.(\ref{single_particle_density}) over $|y|\leq 0.5$.
}\label{Fig:7}
\end{center}
\end{figure}

\begin{table*}[htb!]
\begin{center}
\caption{The values of the freezeout temperatures $T_{\rm F}$ (in
MeV) for different freezeout (F.O.) conditions used and different
values of $q$ investigated.}\label{tab.IIIE3}
\begin{tabular}{|c||c|c|c|c|}\hline
$q$ & \hspace*{5mm} 1.00 \hspace*{5mm}& \hspace*{5mm}1.05 \hspace*{5mm} &
\hspace*{5mm}1.10 \hspace*{5mm}& \hspace*{5mm} 1.15 \hspace*{5mm} \\ \hline\hline
F.O.= $T_{\rm F}$ fixing & 100 & 100 & 100 & 100 \\ \hline
F.O.= $\varepsilon_{\rm F}$ fixing & 100 & 91.8 & 83.2 & 74.3 \\ \hline
F.O.= $s_{\rm F}$ fixing & 100 & 89.3 & 78.5 & 67.4 \\ \hline
\end{tabular}
\end{center}
\end{table*}

\begin{figure}[hbt!]
\begin{center}
\includegraphics [width=8.5cm]{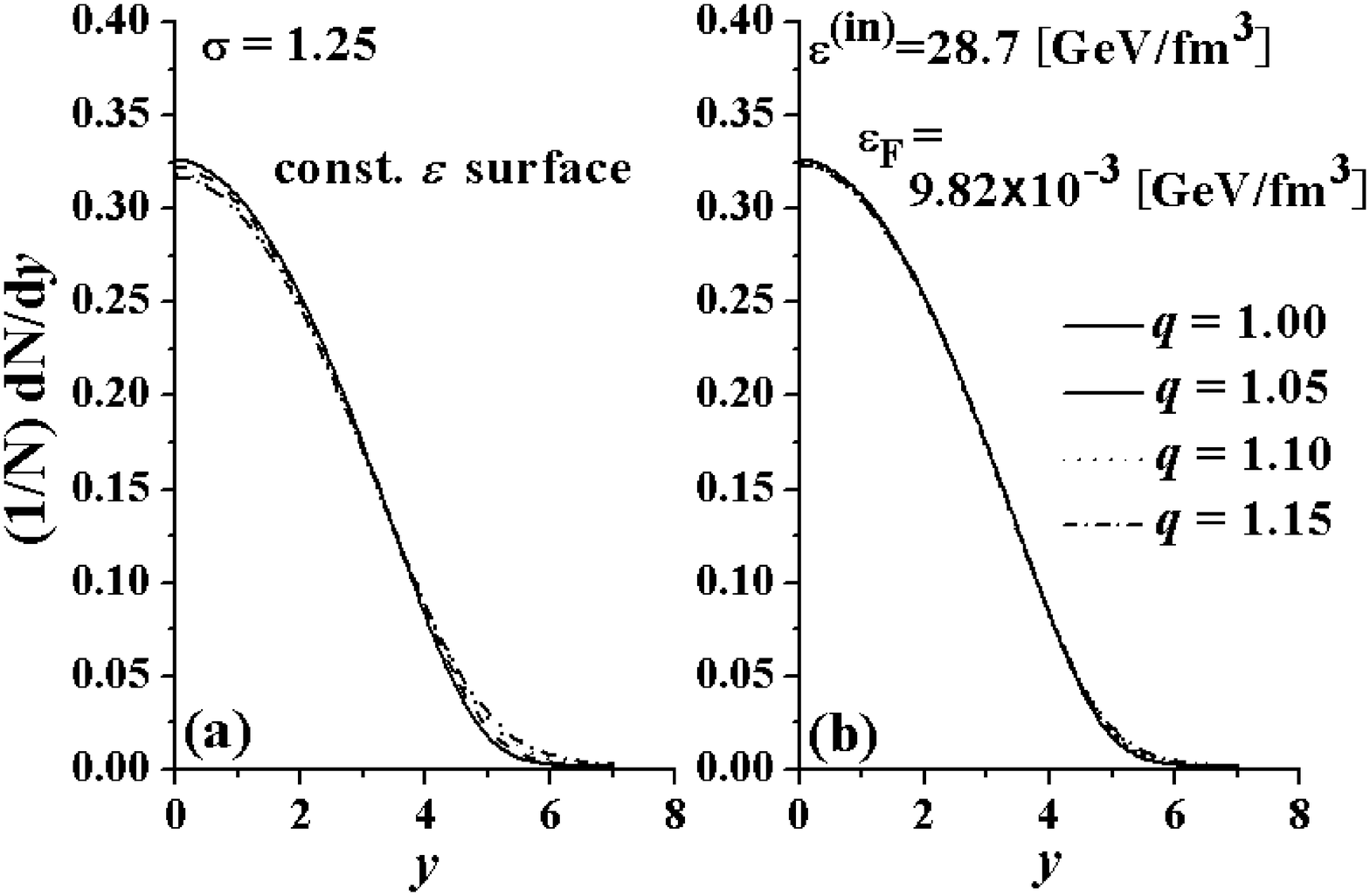}
\includegraphics [width=8.5cm]{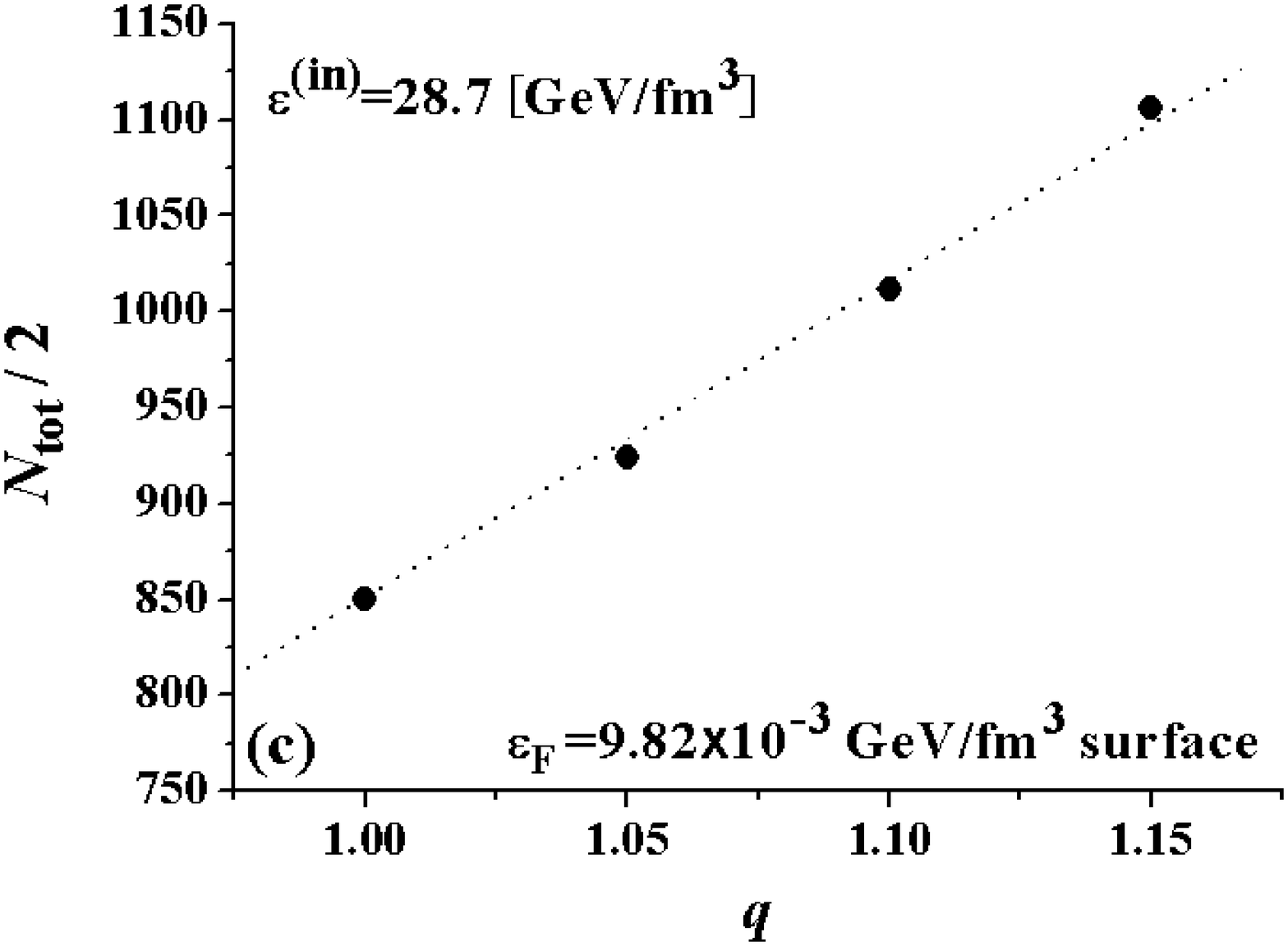}
\caption{ (a) and (b) - normalized rapidity distributions
$1/N~dN/dy$ (defined in the same way is in Fig.~\ref{Fig:7})
calculated for constant energy density freezeout surface
$\varepsilon_{\rm F}$=9.82 $\cdot 10^{-3}$ GeV/fm$^{3}$ for
different values of $q$ using initial conditions with,
respectively, fixed $\sigma = 1.25$ and  fixed $\varepsilon^{\rm (in)}
= 28.7$ GeV/fm$^{3}$. Notice that there is only very weak $q$
dependence confined to small and large regions of rapidity $y$.
(c) - $q$ dependence of the total multiplicity $N_{\rm tot}$
obtained from the $q$-hydrodynamical evolution with fixed
$\varepsilon^{\rm (in)} = 28.7$ GeV/fm$^{3}$ initial condition and
$\varepsilon_{\rm F} = 9.82 \cdot 10^{-3}$ GeV/fm$^{3}$ freezeout
condition. The total multiplicity $N_{\rm tot}$ increases linearly
with $q$. } \label{Fig:8}
\end{center}
\end{figure}

In Fig.~\ref{Fig:7} we show examples of single particle rapidity
and transverse momentum spectra calculated for both types of
initial conditions using $T_{\rm F}$=100~MeV. Note that different
types of the freezeout surface used are connected with using
different sets of parameter  $(q, T_{\rm F})$, cf.
Table.\ref{tab.IIIE3}). Both distributions are sensitive to $q$,
however, in the case of $dN/dy$ this dependence is almost entirely
due to the $q$ dependence of the initial entropy density in the
central region observed in Fig.~\ref{Fig:3} (panels (c) and (f))
and practically vanishes in the case of normalized distributions
calculated for the constant $\varepsilon_{\rm F}$ freezeout
surface, see the panels of Fig.~\ref{Fig:8} (a) and (b).
This is because of the observed $q$ dependence of the corresponding total
multiplicities and is connected with the increase of the entropy
observed in nonextensive processes, see the panel
Fig.~\ref{Fig:8}~(c). We shall discuss this point in more detail below
in Section \ref{sec:meaning}. The weak residual $q$ dependence
observed in this case can be attributed to the (apparently very
weak) effects of the EoS and freezeout surface. In what concerns
$p_{\rm T}$-spectra shown there for different initial conditions
and freezeout surfaces, one observes a very strong dependence on
$q$, which changes the slope of $p_{\rm T}$ considerably. It is
interesting to note that, as seen in Fig.~\ref{Fig:7}, the $p_{\rm
T}$ distributions apparently are not sensitive neither to the type
of initial conditions nor to the freezeout surfaces used.

In the $p_{\rm T}$ distributions, the slope depends on both $q$
and $T_{\rm F}$ and increasing $T_{\rm F}$ while keeping constant
$q$ gives a similar effect as increasing $q$ at fixed $T_{\rm F}$.
On the whole one observes tendency that transverse expansion as
measured by these distributions gets stronger with increasing
nonextensivity, i.e., with increasing $q$.

\section{\label{sec:exp} Comparison with experimental data}

\begin{figure}[hbt!]
\begin{center}
\includegraphics [width=8.5cm]{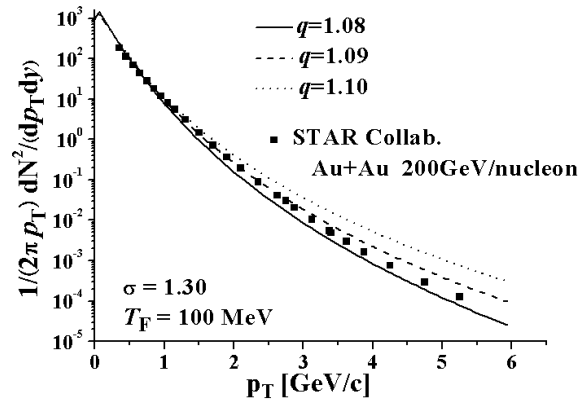}
\caption{Comparison of $q$-hydrodynamic model with experimental
data observed by STAR collaboration\cite{STAR_PhysRevLett97}
performed using $\sigma = 1.30$ and $T_{\rm F} = 100$ MeV for $q =
1.08,~1.09$ and $1.1$ values (with corresponding values of
$\varepsilon^{\rm (in)} = 21.2, 205~$ and $19.7$ GeV/fm$^{3}$). The
best agreement is obtained for $q=1.08$.} \label{Fig:9}
\end{center}
\end{figure}

\begin{figure*}[hbt!]
\begin{center}
\includegraphics [width=12.5cm]{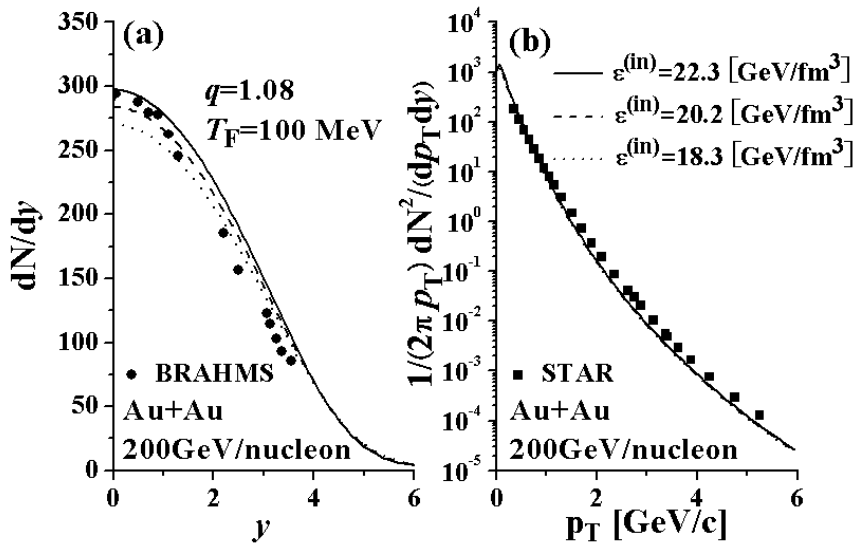}
\caption{Comparison of the $q$-hydrodynamic model with
experimental data on rapidity \cite{BRAHMS_PhysRevLett94} and
$p_{\rm T}$ \cite{STAR_PhysRevLett97} distributions calculated for
$q$=1.08 (fixed) and $T_{\rm F}$=100~MeV, as in Fig. \ref{Fig:9}, but
for different (Gaussian) initial conditions parametrized by
$\varepsilon^{\rm (in)}$.}  \label{Fig:10}
\end{center}
\end{figure*}

We shall confront now our approach with experimental data. Because
of still explanatory character of our work we limit ourselves only
to comparison with some selected rapidity and $p_{\rm T}$ distributions.
At this stage no attempts for exact fits have been made. They must
wait for a more detailed version which, for example, would account
for the possible changes of the nonextensivity parameter $q$
during the collision process as mentioned in Section \ref{sec:I}. 
The same remarks apply to the potentially
promising analysis of anisotropic flow or particle interferometry
(for example, in the way as it was done in
\cite{HiranoPhysRevC66,TeaneyPhysRevC68,CH}), which we postpone
until $1+2$ dimensional version of our approach accounting for
expansion in transverse directions will be available in the
future). Because, as was shown in Section \ref{sec:freeze}, the
most sensitive for $q$ dependence are $p_{\rm T}$ distributions, we
start with them and show in Fig.~\ref{Fig:9} that data by
\cite{STAR_PhysRevLett97} prefer $q=1.08$  and $T_{\rm F} = 100$ MeV (we
attribute the visible discrepancy at largest values of $p_{\rm T}$ to
the contamination from quark jets which carry large momentum in
the initial stage of nuclear collisions and which are not
accounted for $q$-hydrodynamical model. With these  values of $q$
and $T_{\rm F}$ data provided by \cite{BRAHMS_PhysRevLett94} for $dN/dy$
distributions and by \cite{STAR_PhysRevLett97} for $p_{\rm T}$
where compared with predictions of different initial conditions
characterized by $\varepsilon^{\rm (in)}$, see Fig.~\ref{Fig:10}.
As one can see, the $q$-hydrodynamical model with $q$-Gaussian
initial condition can reproduce reasonably well both the rapidity
and transverse momentum distribution data simultaneously.
It should be stressed at this point that with the parameter $q>1$,
which according to the general philosophy of the nonextensive
approach accounts for all possible intrinsic fluctuations in the
system \cite{fluct,Biro,SR}, our model also accounts for the
possible presence of resonances \cite{E2,E3} which therefore, to
avoid double counting, should not be added independently. It must
be noticed that in the present version we do not, in fact, account
for the possible creation of a quark-gluon plasma (QGP) phase.
For this one should use a more elaborate version of EoS than discussed here
in Section \ref{sec:EoS}. Notwithstanding all that, one can say
that a simple $q$-hydrodynamical model reproduces experimental
data reasonably well using $\varepsilon^{\rm (in)} = 19.0\div 22.3$
GeV/fm$^{3}$ ($\sigma= 1.28\div 1.32$), $T_{\rm F} = 100\div 120$
MeV, and $q =1.07\div 1.08$.

\section{\label{sec:meaning}
Discussion: can perfect $q$-hydrodynamics mimic $d$-hydrodynamics?
}

\subsection{Nonextensive/dissipative correspondence - formulation}

Our starting point is observation made at the end of Section
\ref{sec:Examples_of_evolution} that $q$-fluid evolves more slowly
than an ideal fluid. To this one can add observation from Section
\ref{sec:freeze} above that transverse expansion measured by the
behavior of $p_{\rm T}$-spectra is much stronger in $q$-fluid.
Those are precisely features observed in viscous fluids (cf., for
example, \cite{CH}). Let us then treat these observation seriously
and look more closely for the possible connections between
$q$-fluid and viscous fluid apparently emerging from our
$q$-hydrodynamical model.

Let us start with reminding the possible physical meaning of
perfect $q$-hydrodynamics. It originated from the modified
Boltzmann kinetic equation (\ref{q-Boltzmann}) in which a new,
$q$-generalized, version of the Boltzmann molecular chaos
hypothesis \cite{Lavagno,Lima,Biro,BiroK,BiroPurcsel,SR} has been
used in the form of Eq. (\ref{q-molec-chaos}). It can be
introduced in different ways 
\footnote{
                     For example, in \cite{SR} it was a random distortions of energy
                      and momentum conservation caused by the
                      surrounding system which resulted in emergence
                      of some nonextensive equilibrium. In \cite{Biro,BiroK}
                      the two-body energy composition is replaced  by
                      generalized energy sum $h(E_1,E_2)$ (assumed to be
                      associative) which is not necessarily the
                      simple addition and which contains contributions stemming
                      from the pair interaction. It turns out that
                      under quite general assumptions about the function
                      $h$ the division of the total energy
                      among free particles can be done. Different
                      forms of function $h$ lead to different
                      forms of entropy formula, among which one
                      encounters the known Tsallis form as well.
                      The origin of this kind of thinking can be
                      traced to analysis of the $q$-Hagedorn model
                      in \cite{S0}.
} but effectively it
always amounts to postulating a new kind of equilibrium, which
includes some interactions and in which some stationary state is
formed \cite{BiroPurcsel} summarily characterized by parameter
$q$. In our case it leads to Eq.~(\ref{eq:important}), which is
formally identical to perfect hydrodynamical equation but with all
usual ingredients replaced by their $q$-counterparts (perfect
means here that there is nothing on the r.h.s. of
Eq.~(\ref{eq:important})). It is natural to ask how
Eq.~(\ref{eq:important}) would look like when written in terms of
the usual perfect hydrodynamic (with $q=1$) and some reminder
depending on the parameter $q$. Because, as we have seen, in
general $q$ differs only slightly from unity, $q-1 \ll 1$, it is
tempting to simply expand Eq.~(\ref{eq:important}) in small
parameter $|q-1|$ \cite{E1b,S0}. However, as shown in Appendix
\ref{sec:no-approx}, in such case one faces some unsurmountable
problems because terms multiplying $|q-1|$ are not small enough in
the whole of phase space. We shall therefore follow more general
approach.

All our results presented above come from the Eq.
(\ref{eq:important}), which is equation for perfect
$q$-hydrodynamics. Notice that nonextensivity affects not only the
thermodynamical quantities like energy density $\varepsilon$ and
pressure $P$ but also the flow velocity field $u^{\mu}(x)$:
\begin{eqnarray}
\varepsilon(T) &\to& \varepsilon_q(T_q) ~\equiv \varepsilon(T_q)+\Delta\varepsilon_q(T_q),\nonumber\\
P(T) &\to& P_q(T_q) \equiv P(T_q)+\Delta P_q (T_q), \nonumber\\
u^{\mu}(x) &\to& u^{\mu}_q(x) ~\equiv u^{\mu}(x)+\delta u^{\mu}_q(x). \nonumber
\end{eqnarray}
where $u^{\mu}(x)$ is formally a solution of the equation which
has form of dissipative hydrodynamical equation\cite{
Eckart1940,IsraelAnnPhys118,MurongaPRC69,
BaierPhysRevC73,Tsumura2007,HiscockPhysRevD31,ChaudhuriPRC74,
HeinzPRC73,KoidePRC75,DumitruPRC76},
\begin{eqnarray}
\left[\tilde{ \varepsilon} u^{\mu}u^{\nu}
 -\tilde{P} \Delta^{\mu\nu}
 + 2W^{(\mu}u^{\nu)}
 +\pi^{\mu\nu} \right]_{;\mu} \!\!\!=0 .
\label{eq:decomposition}
\end{eqnarray}
The notation used is:
\begin{subequations}
\label{eq:tilde_quantities}
\begin{eqnarray}
&& \tilde{\varepsilon}= 
\varepsilon_q+3\Pi ,
\quad \tilde{P}
=P_q +\Pi,  \label{eq:tilde_q12}\\
&&{W}^{\mu}  
=w_q[1+\gamma] ~\Delta^{\mu}_{\lambda} \delta u_q^{\lambda}, \label{eq:tilde_q3} \\
&& {\pi}^{\mu\nu}  
=\frac{W^{\mu} W^{\nu}}{ w_q [1+\gamma] ^2} +\Pi\Delta^{\mu\nu}  \nonumber \\
&& \hspace*{7mm}=w_q~ \delta u_q^{<\mu}  \delta u_q^{\nu >}
 \label{eq:tilde_q4}
\end{eqnarray}
\end{subequations}
where $\tilde{\varepsilon}$ is energy density, $\tilde{P}$
pressure, $W^{\mu}$  energy or heat flow vector, $\pi^{\mu\nu}$
the (symmetric and traceless) shear pressure tensor and where
\begin{eqnarray}
w_q &\equiv& \varepsilon_q + P_q, \\
\gamma &\equiv& u_{\mu}\delta u_q^{\mu} = -\frac{1}{2}\delta u_{q\mu}\delta u_q^{\mu},
\end{eqnarray}
and
\begin{eqnarray}
A^{( \mu}B^{\nu )} &\equiv& \frac{1}{2}(A^{\mu}B^{\nu} + A^{\nu}B^{\mu}),\nonumber\\
a^{<\mu}b^{ \nu >} &\equiv& [\frac{1}{2}( \Delta^{\mu}_{\lambda}
\Delta^{\nu}_{\sigma} + \Delta^{\mu}_{\sigma}
\Delta^{\nu}_{\lambda} ) -\frac{1}{3}\Delta^{\mu\nu}
\Delta_{\lambda\sigma} ] a^{\lambda}b^{\sigma}\nonumber
\end{eqnarray}
 whereas
\begin{eqnarray}
\Pi  \equiv \frac{1}{3} w_q [\gamma^2+2\gamma] .\label{eq:bulk_pressure}
\end{eqnarray}
This last quantity can be regarded as a bulk pressure to be used
below.

Now comes crucial point of our argumentation. To proceed further
we shall assume that there exists some temperature $T$ and
velocity field $\delta u_q^{\mu}$ satisfying the following
relations:
\begin{subequations}
\label{eq:Nex/diss}
\begin{eqnarray}
&& P(T)= P_q(T_q) \\
&& \varepsilon(T)= \varepsilon_q(T_q) +3\Pi
\end{eqnarray}
\end{subequations}
($\varepsilon$ and $P$ are energy density and pressure defined in
the usual Boltzmann-Gibbs statistics, i.e., for $q=1$). In this
case one can transform equation (\ref{eq:decomposition}) into the
following equation,
\begin{eqnarray}
\left[ \varepsilon(T) u^{\mu}u^{\nu} \!-\!(P(T)+\Pi )
\Delta^{\mu\nu} \!\!\!+\! 2 W^{(\mu} u^{\nu )} \!+\!\pi^{\mu\nu}
\right]_{;\mu} \!\!\!\!=0, \label{eq:Nex/diss_equation}
\end{eqnarray}
which has the familiar form of the usual {\it $d$-hydrodynamical}
equation. This means that perfect $q$-hydrodynamics represented by
Eq.(\ref{eq:important}) can be regarded as being formally {\it
equivalent} to some form of $d$-hydrodynamics as represented by
Eq. (\ref{eq:Nex/diss_equation}). We shall call this observation
the NexDC correspondence (and, respectively, we shall call
Eq.~(\ref{eq:Nex/diss}) with Eq.~(\ref{eq:tilde_q3}) and
(\ref{eq:tilde_q4}) the NexDC relations). This observation can be
traced back to the fact of generic non-conservation of global
entropy in nonextensive systems, cf. Eq. (\ref{eq:S1+S2}),
visualized in Fig. \ref{Fig:8} as increase of the multiplicity
with increasing $q$.

\subsection{Nonextensive/dissipative correspondence - consequences}

We shall now present shortly the most specific immediate
consequences of NexDC correspondence: the entropy production and
estimations of the corresponding transport coefficients.

\subsubsection{Entropy production in $q$-hydrodynamics}

Let us start with observation that Eq.
(\ref{eq:bulk_pressure}) and NexDC relations (\ref{eq:Nex/diss})
lead following form of $q$-enthalpy,
\begin{eqnarray}
 \varepsilon_q(T_q)+P_q(T_q) =\frac{\varepsilon(T)+P(T)}
 {[1+\gamma]^2},
 \label{eq:enthalpy_correspondence}
\end{eqnarray}
which can be also used in definition of $\gamma$ because $w\equiv
Ts=\varepsilon+P$ and $1/(\gamma+1)= \sqrt{1-3\Pi/w}$ ($s$ is the
entropy density in the usual Boltzmann-Gibbs statistics).
Notice that in NexDC one has that
\begin{subequations}
\label{eq:tensor_relations}
\begin{eqnarray}
&& W^{\mu} W_{\mu} \!= -3\Pi w, \\
&& \pi^{\mu\nu}W_{\nu} \!= -2 \Pi W^{\mu}, \\
&& \pi_{\mu\nu}\pi^{\mu\nu} \!= 6\Pi^2.
\end{eqnarray}
\end{subequations}
Suppose now that we define a {\it true equilibrium state} as
a state with $q=1$, i.e., with no residual correlations between
fluid elements and no intrinsic fluctuations present, with energy
momentum tensor
\begin{eqnarray}
{\cal T}^{\mu\nu}_{\rm eq} \equiv {\cal T}^{\mu\nu} = \varepsilon(T)
 u^{\mu}u^{\nu} -P(T)\Delta^{\mu\nu}
\end{eqnarray}
 and with equilibrium distribution
 given by the usual Boltzmann distribution,
\begin{eqnarray}
 f_{\rm eq}(x,p) = \exp \left[-\frac{p^{\mu}u_{\mu}(x)}{k_{\rm B} T(x)}
\right].
\end{eqnarray}
In this case the state characterized by $f_q(x,p)$ given by
Eq.~(\ref{eq:fq}) must be regarded as some stationary state
existing {\it near equilibrium}. Therefore, because we expect that
$|q-1|$ is small, we can define a {\it near equilibrium state}
defined by the correlation function $h_q$ in
Eq.~(\ref{q-molec-chaos}) for which the energy momentum tensor is
 ${\cal T}^{\mu\nu}_q \equiv (\varepsilon_q+P_q)
u^{\mu}_q u^{\nu}_q-P_q g^{\mu\nu}$, c.f.,  Eq.~(\ref{eq:T_q}). It
means then that we can write
\begin{eqnarray}
{\cal T}^{\mu\nu}_q = {\cal T}^{\mu\nu}_{\rm eq} + \delta {\cal T}^{\mu\nu},
\label{eq:T_decomposition}
\end{eqnarray}
where
\begin{eqnarray}
 \delta {\cal T}^{\mu\nu} = -\Pi\Delta^{\mu\nu} +W^{\mu}u^{\nu} +W^{\nu}u^{\mu}  +\pi^{\mu\nu}.
 \label{eq:deltaT}
\end{eqnarray}
Using now Eq. (\ref{eq:Nex/diss}) we obtain the relation
\begin{subequations}
\begin{eqnarray}
\gamma = \sqrt{1+\delta\epsilon_q} -1,
\end{eqnarray}
where
\begin{eqnarray}
\delta\epsilon_q\equiv \frac{\varepsilon(T)-\varepsilon_q(T_q)}{
\varepsilon_q(T_q)+P_q(T_q) },
\end{eqnarray}
\end{subequations}
which connects the velocity field $u$ (solution of the dissipative
hydrodynamics given by Eq.~(\ref{eq:Nex/diss_equation})) with the
velocity field $u_q$ (solution of the $q$-hydrodynamics given by
Eq.~(\ref{eq:important})).

\begin{figure}[hbt!]
\begin{center}
\includegraphics[width=8.5cm]{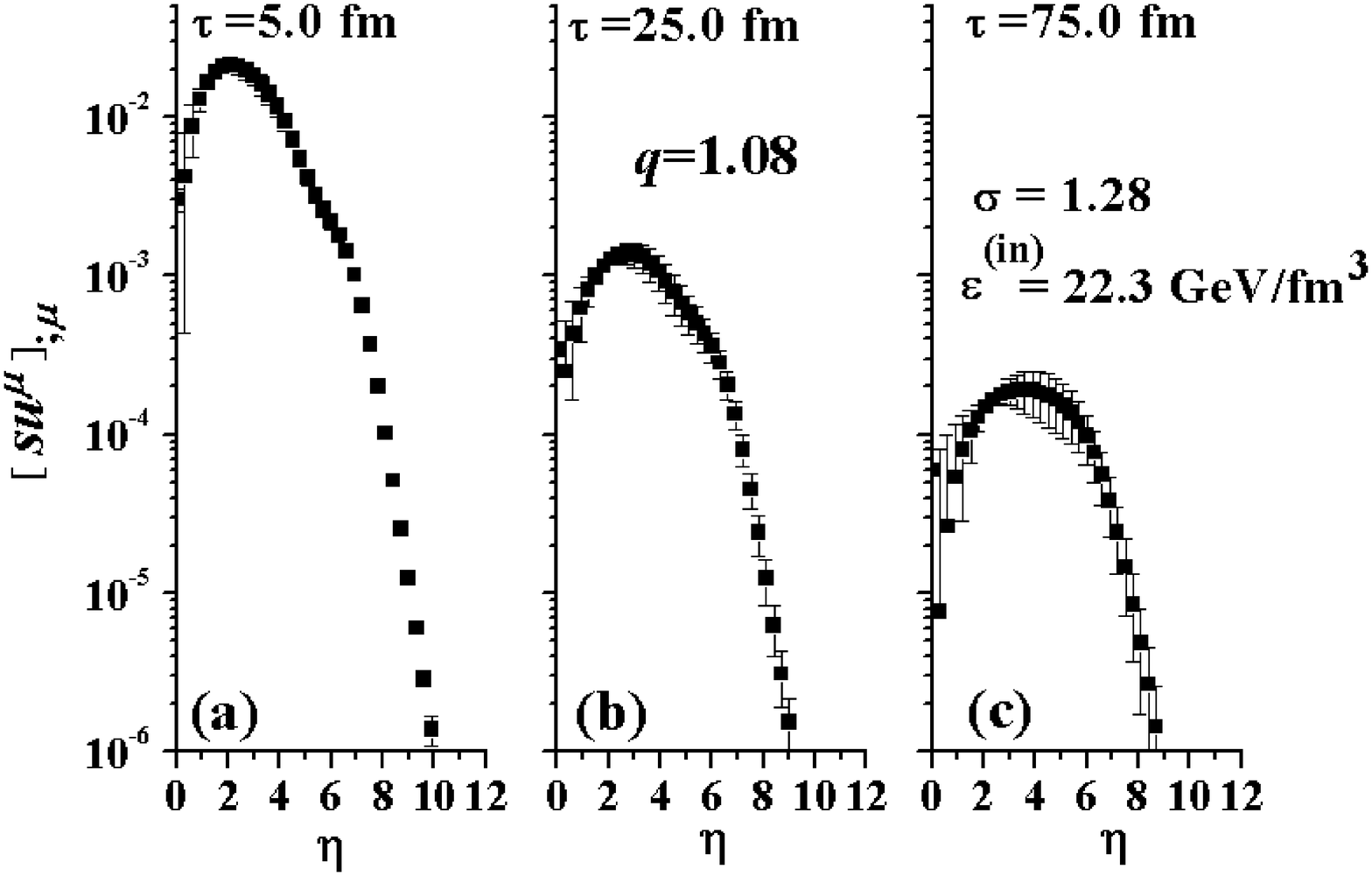}
\includegraphics[width=8.5cm]{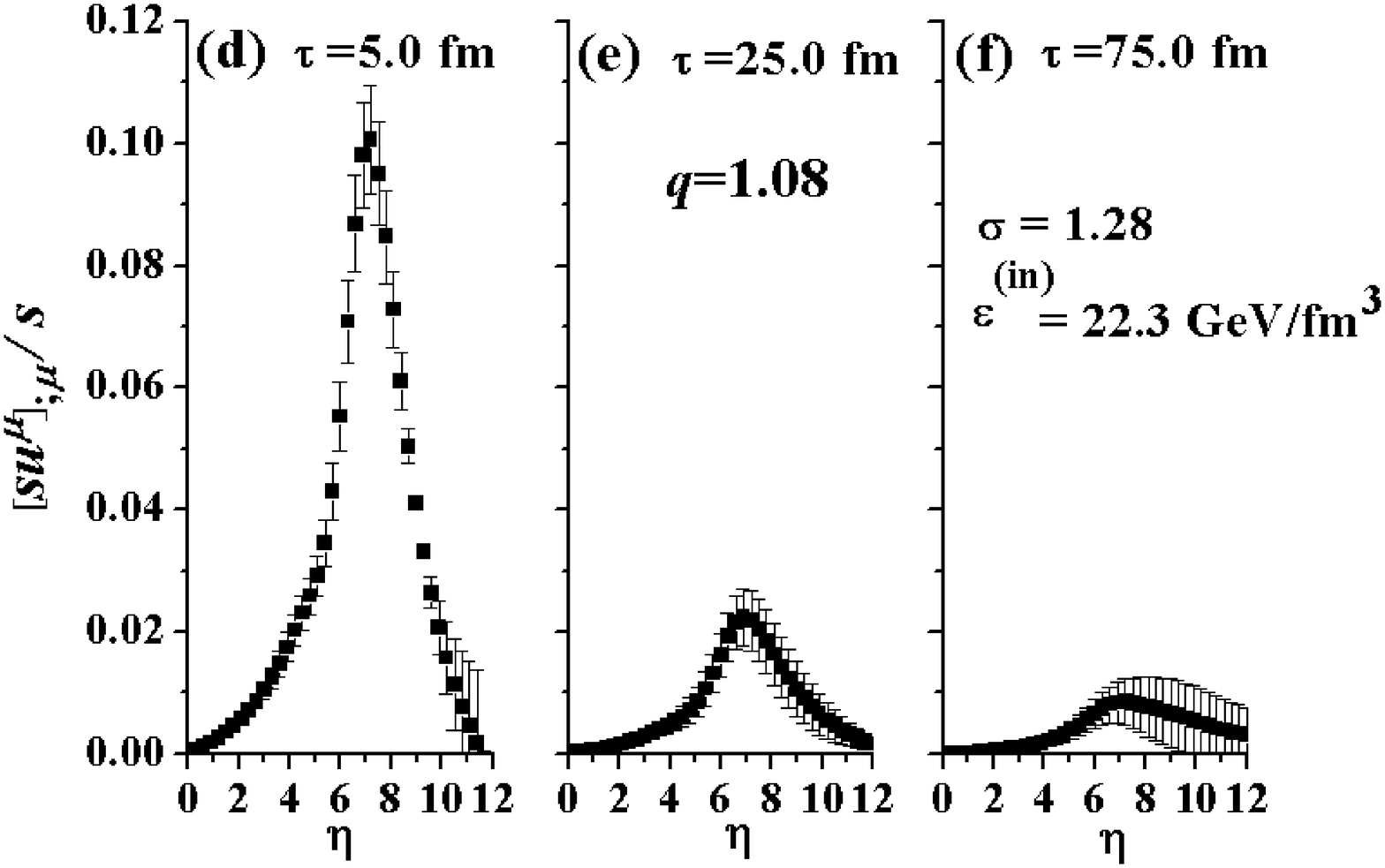}
\caption{
Evolution of the entropy production $[su^{\mu}]_{;\mu}$
(panels (a)-(c)) and ratio $[su^{\mu}]_{;\mu}/s$ (panels
(d)-(f)) as function of $\eta$ for different values of $\tau$
for $q = 1.08$ with $\varepsilon^{\rm (in)} = 22.3$ GeV/fm$^{3}$ (or
$\sigma = 1.28$). The error bar is estimated by the value of
$[s_qu_q^{\mu}]_{\mu}$ obtained in the numerical calculation which
should be zero in an analytical calculation.
}\label{Fig:11}
\end{center}
\end{figure}

In the $(1+1)$ dimensional case discussed here, one can always
parameterize these velocity fields by using the respective fluid
rapidities $\alpha_q$ and $\alpha$, $u_q^{\mu}(x)=\left[
\cosh\left( \alpha_q-\eta \right),\frac{1}{\tau}\sinh\left(
\alpha_q-\eta \right) \right]$ and $u^{\mu}(x)=\left[ \cosh\left(
\alpha-\eta \right),\frac{1}{\tau}\sinh\left( \alpha-\eta \right)
\right]$. Because $\gamma= u_{\mu}\delta
u_q^{\mu}=\cosh(\alpha_q-\alpha)-1$ one has that
\begin{eqnarray}
\cosh(\alpha_q-\alpha) = \sqrt{1+ \delta\epsilon_q},
\label{rapidity_condition1}
\end{eqnarray}
which provides us with a connection between $u$ and $u_q$. From
Eq.~(\ref{rapidity_condition1}), one obtains finally
\begin{eqnarray}
\alpha= \alpha_q- \log\left(  \epsilon_q  + \sqrt{1+\delta\epsilon_q} \right).
\end{eqnarray}
We abandon here another solution of
Eq.~(\ref{rapidity_condition1}), namely that $\alpha= \alpha_q+
\log\left( \epsilon_q  + \sqrt{1+\delta\epsilon_q}
\right)$, because it leads to the entropy reduction, i.e., for it
$[su^{\mu}]_{;\mu} < 0$, for $q>1$. Taking the covariant
derivative of Eq.~(\ref{eq:T_decomposition}) and multiplying it by
$u_{\nu}$ we obtain
\begin{eqnarray}
  u_{\nu}{\cal T}^{\mu\nu}_{q;\mu}=
  T[su_{\mu}]_{;\mu} + u_{\nu} \delta {\cal T}_{;\mu}^{\mu\nu} =0.
\end{eqnarray}
Therefore, although in ideal $q$-hydrodynamics the $q$-entropy is {\it conserved},
i.e., $[ s_q u_q^{\mu}]_{;\mu}=0$, 
we can rewrite it in the form corresponding to dissipative fluid with
{\it entropy production},
\begin{eqnarray}
 [s u^{\mu} ]_{;\mu} =-\frac{u_{\nu}}{T}\delta {\cal T}^{\mu\nu}_{;\mu} .
 \label{eq:sprod_picture1}
\end{eqnarray}
To illustrate this we show in Fig.~\ref{Fig:11} the expected
entropy production as given by Eq.~(\ref{eq:sprod_picture1}).
Notice that $su^{\mu}_{;\mu} > 0$ for the large $\eta$ region at
any $\tau$ (but especially for the early stage of the
hydrodynamical evolution).  It supports therefore a dissipative
analogy of the $q$-hydrodynamics mentioned before and leads us to
very interesting conclusion that equilibrium state which is
generated in the high-energy heavy-ion collisions may in fact be
the $q$-equilibrium state which can be regarded as some stationary
state near the usual (i.e., $q$=1) equilibrium state and which
contains also some dissipative phenomena.

\subsubsection{Calculation of transport coefficients from
$q$-hydrodynamics}

There are different formulations of $d$-hydrodynamics
\cite{Eckart1940,IsraelAnnPhys118,MurongaPRC69,
BaierPhysRevC73,Tsumura2007,HiscockPhysRevD31,ChaudhuriPRC74,
HeinzPRC73,KoidePRC75,DumitruPRC76}. In what follows we shall
choose for comparison only the $2nd$ order theory of dissipative
fluids (in particular as given by \cite{MurongaPRC69,
IsraelAnnPhys118}) leaving investigations of other approaches from
nonextensive perspectives for future investigations. As is known,
this theory does not violate causality (at least not violate the
global causality over distant scale given by the relaxation time),
on the other hand it contains now some dissipative fluxes like
heat conductivity, bulk and shear viscosity. We shall now see to
what extent these transport coefficients can be calculated in
$q$-hydrodynamics.

To this end let us start from considering more closely respective
entropies. Dissipation is connected with the production of entropy
and in \cite{MurongaPRC69, IsraelAnnPhys118} the most general
off-equilibrium four-entropy current $\sigma^{\mu}$ can be written as
\begin{eqnarray}
 \sigma^{\mu} = P(T)\beta^{\mu}
 + \beta_{\nu} ( {\cal T}^{\mu\nu}_{\rm eq} +\delta {\cal T}^{\mu\nu} ) +Q^{\mu},
  \label{eq:s_Israel}
\end{eqnarray}
where $\beta^{\mu}\equiv u^{\mu}/T$ and
$Q^{\mu}=Q^{\mu}\left(\delta \cal{T}^{\mu \nu}\right)$ is some
function which characterizes the off-equilibrium state. In the
case of the $q$-entropy current (\ref{eq:q_entropy}) the NexDC
conjecture (i.e., Eqs.~(\ref{eq:tilde_q3}) and
(\ref{eq:enthalpy_correspondence})) leads to the following
off-equilibrium state:
\begin{eqnarray}
 Q^{\mu} \!\!&=&\!\! Q^{\mu}_{\chi}\equiv
 \chi \left[ su^{\mu} + \frac{W^{\mu}}{T}  \right],
 \label{eq:Q_expression}
\end{eqnarray}
(with $ \chi \equiv \frac{T}{T_q} \sqrt{1-\frac{3\Pi}{w}} - 1$)
which results in
\begin{eqnarray}
\sigma_q^{\mu} ~(\equiv s_q^{\mu}) = ~
 su^{\mu}  + \frac{W^{\mu}}{T}
+\chi \left\{  su^{\mu}  + \frac{W^{\mu}}{T}  \right\} .
\end{eqnarray}
Notice that, because of the strict $q$-entropy conservation
assumed here, using $Q^{\mu}=Q^{\mu}_{\chi}$  one always gets
$\sigma^{\mu}_{q;\mu}=0$. It means that, although there is no
production of $q$-entropy, there is production of the usual
entropy, i.e., our $q$-system is really {\it dissipative} in the
usual meaning of this word.

Let us now be more specific and  use the most general
algebraic form of $Q^{\mu}$, calculated up to the second order in
the dissipative flux, as given by~\cite{MurongaPRC69}
\begin{eqnarray}
  Q^{\mu}_{\rm 2nd} \!\!&=&\!\!
  \frac{\left[ -\beta_0 \Pi^2
                  +\beta_1 W_{\nu}W^{\nu}
                  -\beta_2 \pi_{\nu\lambda}\pi^{\nu\lambda} \right]}  {2T} u^{\mu} \nonumber \\
 &&\hspace*{10mm} - \frac{\alpha_0\Pi W^{\mu}}{T} +
 \frac{\alpha_1\pi^{\mu\nu}W_{\nu}}{T}.
\label{eq:Q_2nd}
\end{eqnarray}
Here $\beta_{i=1,2,3}$ are the corresponding thermodynamic
coefficients for the, respectively, scalar, vector and tensor
dissipative contributions to the entropy current whereas
$\alpha_{i=0,1}$ are the corresponding viscous/heat coupling
coefficients. The $\Pi$ is bulk pressure defined before in Eq.
(\ref{eq:bulk_pressure}) 
\footnote{
               Notice that whereas the time evolution of $\Pi$ is controlled by
                $q$-hydrodynamics (via the respective time dependencies of
                $\varepsilon_q$, $P_q$ and $x$) its form is determined by the
                assumed constraints which must assure that the local entropy
                production in the standard $2^{nd}$ order hydrodynamic theory
                \cite{IsraelAnnPhys118,MurongaPRC69} is never negative. 
}. 
In the NexDC one has that
\begin{eqnarray}
  Q^{\mu}_{\rm 2nd}  \!\! &\to& \!\!
   \Gamma_{\rm 2nd}  ~su^{\mu}  +\Upsilon_{\rm 1st} \frac{W^{\mu}}{T},
\end{eqnarray} where
\begin{subequations}
\label{eqs:2nd_order_theory}
 \begin{eqnarray}
 \Gamma_{\rm 2nd} \!\!&\equiv&\!\!  -\frac{3\beta_1}{2}\Pi  -\frac{(\beta_0+6\beta_2)}{2w} \Pi^2 ,  \\
 \Upsilon_{\rm 1st} \!\!&\equiv&\!\!   - (\alpha_0\ +2 \alpha_1) \Pi  .
 \end{eqnarray}
\end{subequations}
$Q^{\mu}$ can be then expressed by polynomials in the bulk
pressure $\Pi$ defined by Eq. (\ref{eq:bulk_pressure}). It is then
natural to expect that the most general entropy current in the
NexDC approach has form:
\begin{eqnarray}
 Q^{\mu}_{\rm full} = \Gamma(\Pi) su^{\mu} +\Upsilon(\Pi)
 \frac{W^{\mu}}{T},
\end{eqnarray}
where $\Gamma ,\Upsilon$ are (in general infinite) series in
powers of the bulk pressure $\Pi$. In this sense the $Q^{\mu}_{\rm
full}$ can be regarded as the {\it full order dissipative
current}.

In general one has entropy production/reduction,
$\sigma^{\mu}_{;\mu} \ne 0$, however in case when $\Gamma(\Pi)=
\Upsilon(\Pi)=\chi$ one has $\sigma^{\mu}_{\chi;\mu}=0$ so one can
write the full order dissipative entropy current as
\begin{eqnarray}
Q^{\mu}_{\rm full} =
( \chi + \xi )su^{\mu} +
( \chi - \xi )\frac{W^{\mu}}{T},
\end{eqnarray}
where $\Gamma$ and $\Upsilon$ are determined by $\chi\equiv
(\Gamma +\Upsilon)/2$ and $\xi\equiv (\Gamma-\Upsilon)/2$.
From two solution for $(\Gamma,\Upsilon)$,
\begin{subequations}
\label{Q_general}
\begin{eqnarray}
 &&\frac{\Gamma}{2} \equiv \frac{T}{T_q}
\left(\sqrt{1-\frac{3\Pi}{w}}-1 \right),
    \quad \frac{\Upsilon}{2} \equiv \frac{T-T_q}{T_q}  \label{eq:Q_general1}\\
&&\hspace*{-12mm} \mbox{or} \nonumber \\
 && \frac{\Gamma}{2} \equiv  \frac{T-T_q}{T_q},
     \quad \frac{\Upsilon}{2} \equiv \frac{T}{T_q}  \left(\sqrt{1-\frac{3\Pi}{w}}-1
     \right),
      \label{eq:Q_general2}
\end{eqnarray}
\end{subequations}
only (\ref{eq:Q_general1}) is acceptable because only for it
$u_{\mu}Q^{\mu}_{\rm full} \le 0$ (i.e., entropy is maximal in the
equilibrium \cite{MurongaPRC69} and this is because $(T-T_q)/T_q$
is always positive for $q\ge 1$). In this way we finally arrive at
the following possible expression for the full order dissipative
entropy current in the NexDC approach:
\begin{eqnarray}
\sigma^{\mu}_{\rm full} \!\!&\equiv&\!\!
 su^{\mu} +\frac{W^{\mu}}{T} - \nonumber \\
&&\hspace*{-8mm}
-\frac{2T}{T_q}\left[~1- \sqrt{1-\frac{3\Pi}{w}} ~\right]su^{\mu} +
\frac{2(T-T_q)}{T_q}\frac{W^{\mu}}{T}.\quad
\label{eq:full_order_current}
\end{eqnarray}
Limiting ourselves to situations when $T/T_q \approx 1$ and
neglecting terms higher than ${\cal O}(3\Pi/w)^2$, one obtains
that
\begin{eqnarray}
Q^{\mu}_{\rm full}  \!\!&\approx&\!\!
\left[ -\left( \frac{3\Pi}{w} \right) -\frac{1}{4} \left( \frac{3\Pi}{w} \right)^2
\right] su^{\mu}. \label{eq:Q_chi_2nd}
\end{eqnarray}
Comparing now Eqs.~(\ref{eqs:2nd_order_theory}) and
(\ref{eq:Q_chi_2nd}) one gets that 
\footnote{
                   The fact that we obtained non-zero coefficients
                    $\beta_{i=1,2,3}$ and couplings $\alpha_{i=0,1}$
                    for dissipative flux of Eq.~(\ref{eq:Q_2nd}) as
                    found in Eq.~(\ref{correspondence_in2nd}) means that
                    $d$-hydrodynamics obtained via NexDC from
                    $q$-hydrodynamics accounts for all second order terms.
                    One may conclude than it seems that such $d$-hydrodynamics
                    with full order entropy current has global causality
                    (over a distance scale given by relaxation time).
                    However, question whether NexDC violates causality
                    remains so far unsettled.
}
\begin{eqnarray}
 \beta_1 = \frac{2}{w}, \quad \beta_0+ 6\beta_2 = \frac{9}{2w},
 \quad \alpha_0+2\alpha_1=0.
\label{correspondence_in2nd}
\end{eqnarray}
Since in the Israel-Stewart theory \cite{IsraelAnnPhys118} the
relaxation time $\tau$ is proportional to thermodynamical
coefficients $\beta_{0,1,2}$, it is naturally to assume that in
our NexDC case $\tau \propto 1/w$, i.e., it is proportional to the
inverse of the enthalpy (notice that for classical Boltzmann gas
of massless particles one obtains $\beta_2=3/w$
\cite{MurongaPRC69,DumitruPRC76}).

We shall now derive the bulk and shear viscosities emerging from
the NexDC approach. Let us start with observation that the local
entropy production by the full order entropy current
Eq.~(\ref{eq:full_order_current}) can be also written as
\begin{eqnarray}
\sigma^{\mu}_{{\rm full};\mu} \!\!&=&\!\!
[(1+\chi)\Phi^{\mu}]_{;\mu} + [\xi\Psi^{\mu}]_{;\mu},
\label{eq;sigma_full0}
\end{eqnarray}
where $\Phi^{\mu}=su^{\mu}+\frac{W^{\mu}}{T}$ and $\Psi^{\mu}=
su^{\mu}-\frac{W^{\mu}}{T}$. Because conservation of $q$-entropy,
$\sigma^{\mu}_{q;\mu}=0$, is equivalent to
$[(1+\chi)\Phi^{\mu}]_{;\mu} =0$, therefore using
Eq.~(\ref{eq:enthalpy_correspondence}) one gets that
\begin{eqnarray}
\Psi^{\mu} \!\!&=&\!\! -\frac{W^{\nu}W_{\nu}}{3\Pi T} u^{\mu}+ \frac{ W_{\nu}}{2\Pi T} \pi^{\mu\nu} \label{eq:PhiPsi}
\end{eqnarray}
and (see Appendix \ref{sec:shear_visco} for details of derivation
of Eqs. (\ref{eq:sigma_full1}) and (\ref{eq:sigma_full2}))
\begin{eqnarray}
\sigma^{\mu}_{{\rm full};\mu} \! = \! -\frac{\Pi}{T}
(wu^{\mu}X_{\mu}) -\frac{W^{\mu}}{T} \tilde{Y}_{\mu} +
\frac{\pi^{\mu\nu}}{T} Z_{\mu\nu},  \label{eq:sigma_full1}
\end{eqnarray}
where
\begin{subequations}
\begin{eqnarray}
X_{\mu} \!\!&=&\!\! -\frac{\xi}{\Pi} \left[
\frac{\partial_{\mu}\Pi}{\Pi} +\frac{\partial_{\mu} T}{T}
+\frac{\partial_{\mu} \xi}{\xi}
\right],\\
Y_{\mu} \!\!&=&\!\! \frac{\xi}{\Pi} \left[ \frac{2}{3} u^{\nu}W_{\mu;\nu} +\frac{1}{3}W_{\mu}u^{\nu}_{;\nu} -\frac{1}{2}\pi^{\nu}_{\mu;\nu} \right],\\
Z_{\mu\nu} \!\!&=&\!\! \frac{\xi}{\Pi}\left[  \frac{1}{2}
W_{\nu;\mu} \right]
\end{eqnarray}
and
\begin{eqnarray}
\tilde{Y}_{\mu}\!\! &=&\!\! Y_{\mu} - \Pi X_{\mu}, \quad
\tilde{Z}_{\mu\nu}\equiv Z_{\mu\nu}+ \frac{\tilde{Y}_{\mu}
W_{\nu}}{2\Pi} .
\end{eqnarray}
\end{subequations}
One can now use Eq.~(\ref{eq:tensor_relations}) to eliminate
term proportional to heat flow $\frac{W^{\mu}}{T}$. In this way
one avoids explicit contribution to entropy production from the
heat flow $\frac{W^{\mu}}{T}$, which is present in
Eq.~(\ref{eq:sigma_full1}) when one discuss baryon free fluid, in
which case the necessity to use Landau frame would appear. As one
can see Eq.~(\ref{eq;sigma_full0}) is covariant and therefore it
does not depend on the frame used. After that one obtains that
\begin{eqnarray}
\sigma^{\mu}_{{\rm full};\mu}= -\frac{\Pi}{T} (wu^{\mu}X_{\mu})
+\frac{\pi^{\mu\nu}}{T} \tilde{Z}_{\mu\nu}. \label{eq:sigma_full2}
\end{eqnarray}
Eq.~(\ref{eq:sigma_full2}) can be now used to find the bulk and
shear viscosities from $\sigma_{{\rm full};\mu}^{\mu}$ given by
Eq.~(\ref{eq;sigma_full0}). The positive transport coefficients,
bulk viscosity $\zeta$ and shear viscosity $\eta$, can be
estimated by writing the entropy production $\sigma^{\mu}_{{\rm
full};\mu}$ in the following form:
\begin{eqnarray}
\sigma^{\mu}_{{\rm full};\mu} = \frac{\Pi^2}{\zeta T} +
\frac{\pi^{\mu\nu}\pi_{\mu\nu}}{2\eta T} \ge 0
\end{eqnarray}
and using Eq.~(\ref{eq:tensor_relations}). We arrive then at the
sum rule connecting transport coefficients (expressed as ratios of
bulk and shear viscosities over the entropy density $s$),
\begin{eqnarray}
 \frac{1}{\zeta/s}+\frac{3}{\eta/s} = \frac{w \sigma^{\mu}_{{\rm full};\mu}}{\Pi^2}.
\label{eq:sum_rule}
\end{eqnarray}

This is as far as we can go. The heat conductivity, as shown
above, can be expressed by two other transport coefficients for
which we have only one equation in the form of sum rule
(\ref{eq:sum_rule}). To proceed any further and to disentangle
(\ref{eq:sum_rule}) one has to add some additional input. Suppose
then that we are interested in the extremal situation, when total
entropy is generated by action of shear viscosity only. In this
case one can rewrite Eq.~(\ref{eq:sigma_full2}) as
\begin{eqnarray}
 \sigma^{\mu}_{{\rm full};\mu} = \frac{\pi^{\mu\nu}}{T}\left[ - \frac{\pi_{\mu\nu}}{6\Pi} (wu^{\lambda}X_{\lambda})
 +\tilde{Z}_{\mu\nu}  \right],
\end{eqnarray}
resulting in
\begin{eqnarray}
\frac{\eta}{s} = \frac{\gamma(\gamma+2)}{3(\gamma+1)^2}  \left[
\frac{\pi^{\mu\nu}}{\Pi}\frac{\tilde{Z}_{\mu\nu}}{T} -
su^{\lambda} X_{\lambda} \right]^{-1}.
\label{eq:nex/diss_prediction}
\end{eqnarray}
\begin{figure}[hbt!]
\begin{center}
\includegraphics[width=8.5cm]{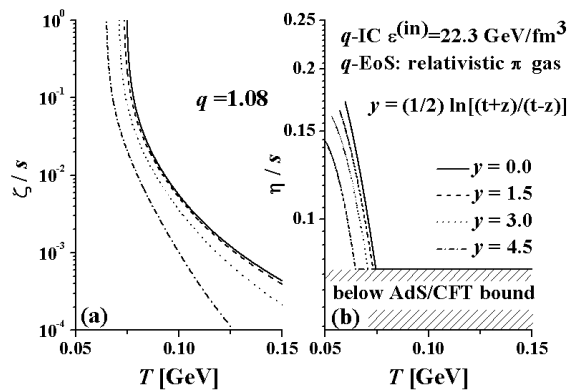}
\caption{The NexDC predictions for the ratios of bulk (panel (a))
and shear  (panel (b)) viscosities over the entropy density,
$\zeta/s$ and $\eta/s$, as function of temperature $T$ and
calculated for a number of space-time rapidities  $y\equiv
\frac{1}{2}\ln\frac{t+z}{t-z}$ (the same as in
Eq.~(\ref{eq:taueta}) using $q$-hydrodynamical
model).}\label{Fig:12}
\end{center}
\end{figure}
\noindent Note that Eq.~(\ref{eq:nex/diss_prediction}) allows all
values of $\eta/s$, in particular that $\eta/s <\frac{1}{4\pi}$,
what violates the limit obtained from AdS/CFT correspondence that
$\eta/s\ge 1/4\pi$ \cite{KovtumPRL94}. To impose this limit we
shall now use Eq.~(\ref{eq:sum_rule}). This can be done only in
the region where r.h.s. of Eq.~(\ref{eq:nex/diss_prediction}) is
smaller than (or equal to) $1/4\pi$ (in which case we put $\eta/s
= 1/4\pi$), otherwise, because of our earlier assumed limitation
we put $\zeta/s=0$ and use Eq.~(\ref{eq:nex/diss_prediction}) to
evaluate $\eta/s$. The corresponding results for $\zeta/s$ and
$\eta/s$ are shown in Fig.\ref{Fig:12} (a) and (b), respectively.
Notice that when the r.h.s of Eq.~(\ref{eq:nex/diss_prediction})
approaches $1/4\pi$, $\zeta/s$ given by Eq.~(\ref{eq:sum_rule})
approaches infinity. All curves presented in Fig. \ref{Fig:12}
were calculated for Au+Au collisions with identical set of
parameter as the best fit presented in Fig. \ref{Fig:10} above.

\section{\label{sec:SumCon}Summary and conclusions}

We have presented a nonextensive version of the hydrodynamical
model for multiparticle production processes, the
$q$-hydrodynamical model, which is based on the nonextensive
statistics represented by Tsallis entropy and indexed by the
nonextensivity parameter $q$. In doing so we have followed the
usual approach originating in the appropriate kinetic equations
formulated in nonextensive form in \cite{Lavagno}. We have found
the nonextensive entropy current which satisfies not only the
nonextensive $H$-theorem, Eq.~(\ref{eq:q_entropy}), but also the
$q$-version of thermodynamical relations
Eq.~(\ref{eq:thermodynamical-relation}). The (1+1) dimensional
$q$-hydrodynamics with the $q$-Gaussian initial condition and the
$q$-EoS can reasonably reproduce the single particle spectra
observed at RHIC energy for $q=1.07\div 1.08$ for $T_{\rm F} =
100\div 120$ MeV  if quark jet contributions to $p_{\rm T}$
spectra are small, i.e., up to transverse momentum range around
$p_{\rm T} \le  6.0$ GeV/c. We also found a possible
correspondence between the $q$-hydrodynamics and the usual ($q=1$)
$d$-hydrodynamics (NexDC) as provided by the
Eq.~(\ref{eq:Nex/diss_equation}) with NexDC relations
Eq.~(\ref{eq:Nex/diss}). Based on this correspondence, we have
evaluated the entropy production in relativistic heavy-ion
collisions at RHIC energy using results of our perfect
$q$-hydrodynamics (understood as an approach without any
$q$-viscosity effects added). The fact that when comparing with
data one finds that $q>1$ means that one, indeed, has some dynamic
factors present, detailed form of which is not yet disclosed but
which summarily can be accounted for by the nonextensive approach
and which action is summarized by the parameter $q-1$.

In what concerns the obtained $p_{\rm T}$ dependence, our formula
continues attempts made to interpret power-law spectra as new kind
of equilibrium phenomena for the whole $p_{\rm T}$ range pushing the
usual interpretation via onset of "hard" collisions (imposed on
the "soft" ones) to really high values of $p_{\rm T}$ (cf.,
\cite{E3,E1b,BiroPurcsel} and references therein). In such
approach there is no characteristic scale at which the transition
from "soft" (or locally thermalized) to "hard" (or unthermalized)
dynamics occurs which appears in the conventional descriptions
using viscous hydrodynamics, as for example in
\cite{TeaneyPhysRevC68,Add1}.

One of the results of our investigation is that fluctuations in
the initial conditions seem to be the most important part of the
hydrodynamical model, which by using Tsallis statistics, attempts
to account for any possible fluctuations in some general, model
independent, way. This is quite reasonable result because at the
initial stage our system consists of a relatively small number of
degrees of freedom and is therefore more sensitive to any
fluctuations. On the contrary, at freezeout this number is much
bigger and the system is only weakly responding to any
fluctuations. This finding agrees nicely with recent analysis of
the elliptical flow performed by using a hydrodynamical approach
in which an attempt was made to account for fluctuations (without,
however, using $q$-statistics), see \cite{ADHKS}. On the other
hand, however, it should be remembered that analysis presented
here is considerably simplified by using the same nonextensivity
parameter $q$ at all stages of the collision process. There is
therefore room for improvements which will facilitate comparison
with data. One can argue that intrinsic fluctuations existing in
different stages of the collision process are of different (albeit
connected) dynamical origin and therefore parameters $q$ for the
initial conditions, for the EoS and, finally, for the
hydrodynamical expansion should be allowed to have different
values (and should be also different for the longitudinal and
transverse dynamics). The other problem would be how to connect
our $q$ parameter expressing fluctuations with fluctuations in all
momentum observables as seen when analyzing a non-ideal liquid as
it was done, for example, in \cite{Vogel}. We plan to address
these problems elsewhere. In any case, similarly to the fact that
concept of ideal fluid is never realized in nature \cite{CH} (the
bound of $\eta/s \ge 1/4\pi$ found in \cite{KovtumPRL94} being
strong argument supporting this), the $q=1$ case should be
replaced by investigations of the $q$-fluid with $q>1$.

In this context the natural question arises concerning the deeper
physical meaning of the $q$-hydrodynamic proposed here. The most
important observation discussed in Section \ref{sec:meaning} is
the apparent correspondence found between the perfect
$q$-hydrodynamics and the usual $d$-hydrodynamics, which we call
the NexDC correspondence. It allows calculation of transport
coefficients of viscous fluid in terms of parameters of
$q$-(ideal) fluid, i.e., essentially as dependent on single
parameter which, as it was already stressed many times, represents
summary effect of many possible dynamical factors, without
entering into dynamical details (i.e., in a purely
phenomenological way). The detail discussion of the NexDC
phenomenon is, however, outside the scope of present paper and we
plan to address is elsewhere.

We close by remark that hydrodynamics can also be derived using
Information Theory with its method of maximization of
information entropy under some specific constraints
\cite{MaxEnthydro}. It is therefore plausible that our results
could also be derived using a nonextensive version of Information
Theory (in the way as \cite{Kq,qMaxEnt} can be regarded as a
nonextensive generalization of the Information Theory approach to
single particle distributions obtained in the multiparticle
production processes proposed in \cite{MaxEnt}). We shall not
pursue this possibility here.\\

\begin{acknowledgments}
One of us (TO)
thanks the Yukawa Institute for Theoretical Physics at Kyoto
University. Discussions during the YITP workshop YITP-W-07-07 on
"Thermal Quantum Field Theories and Their Applications" were
useful to complete this work. Partial support (GW) of the Ministry
of Science and Higher Education under contracts
1P03B02230 and  CERN/88/2006 is acknowledged.
\end{acknowledgments}
%
\appendix

\section{\label{sec:derivation}Derivation of eqs.(\ref{eq:3.21}) and (\ref{eq:3.22})}

Consider some details of the $(1+1)$ dimensional relativistic
hydrodynamics under assumption that one ignores the transverse
expansion of the fluid. With $g_{\mu\nu}={\rm diag}(1,-\tau^2)$ and four
fluid velocity $u_{\mu}=[\cosh(\alpha_q-\eta),
-\tau\sinh(\alpha_q-\eta)]$ the projection matrix is equal to:
\begin{widetext}
\begin{eqnarray}
\Delta_{q \mu\nu}\equiv g_{\mu\nu}-u_{q\mu}u_{q\nu} =
\left[\begin{array}{ll}
-\sinh^2(\alpha_q-\eta)\quad & \tau\cosh(\alpha_q-\eta)\sinh(\alpha_q-\eta) \\
\tau\cosh(\alpha_q-\eta)\sinh(\alpha_q-\eta) & \quad-\tau^2\cosh^2(\alpha_q-\eta)\\
\end{array}
\right].
\end{eqnarray}
\end{widetext}
The nonvanishing components of Christoffel symbols are
$\Gamma^{\eta}_{\tau\eta}=\Gamma^{\eta}_{\eta\tau}=1/\tau$  and
$\Gamma^{\tau}_{\eta\eta}=\tau $, therefore covariant derivative of fluid
velocity, which is defined by
$u^{\mu}_{;\nu}=\partial_{\nu}u_{\mu}+\Gamma_{\lambda\nu}^{\mu}u^{\lambda}$,
has the following form:
\begin{eqnarray*}
&& u^{\tau}_{q;\tau}=\sinh(\alpha_q-\eta)\frac{\partial\alpha_q}{\partial\tau},~
u^{\eta}_{q;\tau}=\frac{1}{\tau}\cosh(\alpha_q-\eta)\frac{\partial\alpha_q}{\partial\tau}, \\
&&
u^{\tau}_{q;\eta}=\sinh(\alpha_q-\eta)\frac{\partial\alpha_q}{\partial\eta},~
u^{\eta}_{q;\eta}=\frac{1}{\tau}\cosh(\alpha_q-\eta)\frac{\partial\alpha_q}{\partial\eta}.
\end{eqnarray*}
Using these expressions, one obtains
\begin{eqnarray*}
 && u_{q}^{\mu}\Delta_{q \tau\nu}u^{\nu}_{q;\mu} =
 ~\cosh(\alpha_q-\eta)\sinh(\alpha_q-\eta)\frac{\partial\alpha_q}{\partial\tau} \\
 &&\hspace*{20mm} +\frac{1}{\tau} \sinh^2(\alpha_q-\eta)\frac{\partial\alpha_q}{\partial\eta}, \\
&&u_{q}^{\mu}\Delta_{q\eta\nu}u^{\nu}_{q;\mu}=
-\tau \cosh^2(\alpha_q-\eta)\frac{\partial\alpha_q}{\partial\tau} \\
&&\hspace{20mm} - \cosh(\alpha_q-\eta)\sinh(\alpha_q-\eta)\frac{\partial\alpha_q}{\partial\eta}.
\end{eqnarray*}
Because $g^{\mu\nu}_{;\nu}=0$ for $\mu,\nu=\tau$ and $\eta$,
Eq.~(\ref{second_equation}) is reduced to the following two
equations:
\begin{eqnarray}
&& (\varepsilon_q+P_q) \left\{ u_{q}^{\mu}\Delta_{q \tau\nu} u^{\nu}_{q;\mu}\right\} \nonumber \\
&& \hspace*{15mm}
-\Delta_{q \tau\tau} \frac{\partial P_q}{\partial \tau}
+\frac{1}{\tau^2} \Delta_{q \tau\eta}\frac{\partial P_q}{\partial\eta} =0,\label{eq:B2} \\
&& (\varepsilon_q+P_q) \left\{ u_{q}^{\mu}\Delta_{q \eta\nu}u^{\nu}_{q;\mu}\right\} \nonumber \\
&& \hspace*{15mm}
-\Delta_{q \eta\tau} \frac{\partial P_q}{\partial \tau}
+\frac{1}{\tau^2} \Delta_{q \eta\eta}\frac{\partial P_q}{\partial\eta} =0\label{eq:B3}.
\end{eqnarray}
The equations (\ref{eq:B2}) and (\ref{eq:B3}) are equivalent, therefore
one has only one equation,
\begin{eqnarray}
 && (\varepsilon_q+P_q) \left\{ \frac{\partial\alpha_q}{\partial\tau}
  +\frac{ \tanh(\alpha_q-\eta)}{\tau} \frac{\partial\alpha_q}{\partial\eta} \right\} \nonumber \\
 && \hspace*{15mm}+ \tanh(\alpha_q-\eta) \frac{\partial P_q}{\partial\tau}
  + \frac{1}{\tau}\frac{\partial P_q}{\partial\eta} =0,
\end{eqnarray}
which is the Eq.~(\ref{eq:3.22}).
Since the four divergence of the fluid velocity $u^{\mu}_{q;\mu}$ is given by
\begin{eqnarray*}
 u^{\mu}_{q;\mu} = \sinh(\alpha_q-\eta)\frac{\partial\alpha_q}{\partial\tau} + \frac{1}{\tau}
 \cosh(\alpha_q-\eta)\frac{\partial\alpha_a}{\partial\eta},
\end{eqnarray*}
the Eq.~(\ref{second_equation}) can be written as
\begin{eqnarray*}
  && \cosh(\alpha-\eta)\frac{\partial\varepsilon_q}{\partial\tau} +
  \frac{\sinh(\alpha_q-\eta)}{\tau}  \frac{\partial\varepsilon_q}{\partial\eta}
  \nonumber \\
&& \hspace*{15mm}+ (\varepsilon_q+P_q) \bigg \{\sinh(\alpha_q-\eta)\frac{\partial\alpha_q}{\partial\tau} \nonumber \\
&& \hspace*{25mm}+\frac{1}{\tau}\cosh(\alpha_q-\eta)\frac{\partial\alpha_a}{\partial\eta}\bigg\} =0,
\end{eqnarray*}
leading immediately to
\begin{eqnarray}
&& \frac{\partial\varepsilon_q}{\partial\tau} +
  \frac{\tanh(\alpha_q-\eta)}{\tau}  \frac{\partial\varepsilon_q}{\partial\eta} \nonumber \\
&& \hspace*{0mm}
 + (\varepsilon_q+P_q) \left\{
  \tanh(\alpha_q-\eta)\frac{\partial\alpha_q}{\partial\tau} + \frac{1}{\tau}
  \frac{\partial\alpha_a}{\partial\eta}
  \right\} =0, \nonumber \\
\end{eqnarray}
which is Eq.~(\ref{eq:3.21}).
%
\begin{widetext}
\section{\label{sec:Numerics}Numerical method used}

For the purpose of numerical calculations we express
Eqs.(\ref{eq:3.21}) and (\ref{eq:3.22}) in the form of finite
difference equations :
\begin{eqnarray}
  &&
  ~~A_{1(j)}^{~(n)} \left\{ \frac{~\varepsilon_{q(j)}^{~(n+1)}
   - \frac{1}{2}[ ~\varepsilon_{q(j+1)}^{~(n)} + \varepsilon_{q(j-1)}^{~(n)}]
     }{\Delta\tau} \right\} +A_{2(j)}^{~(n)} \left\{ \frac{\varepsilon_{q(j+1)}^{~(n)}
  - \varepsilon_{q(j-1)}^{~(n)}}{2\Delta\eta} \right\} ,
 \nonumber \\
  &&
  +A_{3(j)}^{~(n)} \left\{ \frac{\alpha_{q(j)}^{~(n+1)}
   -\frac{1}{2}[ \alpha_{q(j+1)}^{~(n)} + \alpha_{q(j-1)}^{~(n)}]
  }{\Delta\tau} \right\}
 +A_{4(j)}^{~(n)} \left\{ \frac{\alpha_{q(j+1)}^{~(n)}
   - \alpha_{q(j-1)}^{~(n)}}{2\Delta\eta} \right\}=0
\end{eqnarray}
and \begin{eqnarray}
&&  ~~B_{1(j)}^{~(n)} \left\{
\frac{P_{q(j)}^{~(n+1)}
   - \frac{1}{2}[ P_{q(j+1)}^{~(n)}+P_{q(j-1)}^{~(n)}] }{\Delta\tau} \right\}
 +B_{2(j)}^{~(n)} \left\{ \frac{P_{q(j+1)}^{~(n)}
   - P_{q(j-1)}^{~(n)}}{2\Delta\eta} \right\},
 \nonumber \\
&&
+B_{3(j)}^{~(n)} \left\{ \frac{\alpha_{q(j)}^{~(n+1)}
   - \frac{1}{2}[ \alpha_{q(j+1)}^{~(n)} + \alpha_{q(j-1)}^{~(n)}]
  }{\Delta\tau} \right\}
 +B_{4(j)}^{~(n)} \left\{ \frac{\alpha_{q(j+1)}^{~(n)}
   - \alpha_{q(j-1)}^{~(n)}}{2\Delta\eta} \right\}=0 .
\end{eqnarray}
The subscript $(j)$ and superscript $(n)$ represent the
corresponding grid number in the $\eta$ and $\tau$ space with grid
spacings $\Delta\eta$ and $\Delta\tau$, respectively, i.e., with
$\eta_j=j\Delta\eta$ and $\tau_n=\tau_0+n \Delta\tau$. The
coefficients appearing in the above equations are defined in the
following way:
\begin{equation}\begin{array}{ll}
   A_{1(j)}^{~(n)} \equiv 1,
& B_{1(j)}^{~(n)} \equiv v_{q(j)}^{~(n)},   \\
   A_{2(j)}^{~(n)} \equiv [ v_{q(j)}^{~(n)} ]/ {\tau_n},
& B_{2(j)}^{~(n)} \equiv 1/{\tau_n}, \\
   A_{3(j)}^{~(n)} \equiv (\varepsilon_{q(j)}^{~(n)}+P_{q(j)}^{~(n)}) [v_{q(j)}^{~(n)}] ,
& B_{3(j)}^{~(n)} \equiv (\varepsilon_{q(j)}^{~(n)}+P_{q(j)}^{~(n)}) ,\\
   A_{4(j)}^{~(n)} \equiv (\varepsilon_{q(j)}^{~(n)}+P_{q(j)}^{~(n)})/ {\tau_n},
& B_{4(j)}^{~(n)} \equiv (\varepsilon_{q(j)}^{~(n)}+P_{q(j)}^{~(n)})[v_{q(j)}^{~(n)}]/{\tau_n} .
\end{array}
\end{equation}
Introducing now notation,
\begin{eqnarray}
    c_{s~(j)}^{2~(n)}
    \equiv \frac{ P_{q(j)}^{~(n)}} {\varepsilon_{q(j)}^{~(n)}},
\end{eqnarray}
where $c_{s~(j)}^{2~(n)}$ is function of $\varepsilon_{q(j)}^{~(n)}$
(due to the equation of state $P_q(\varepsilon_q)$), one can rewrite
above two equations in the following form:
\begin{eqnarray}
   &&\bigg[ A_{1(j)}^{~(n)} \bigg]\varepsilon_{q(j)}^{~(n+1)}
    +\bigg[ A_{3(j)}^{~(n)} \bigg] \alpha_{q(j)}^{~(n+1)} \hspace*{75mm}
\nonumber \\
   && \hspace*{12mm}
     -\frac{1}{2} \bigg[A_{1(j)}^{~(n)} - A_{2(j)}^{~(n)} \frac{\Delta\tau}{\Delta\eta}
       \bigg] \varepsilon_{q(j+1)}^{~(n)}
     -\frac{1}{2} \bigg[ A_{1(j)}^{~(n)} + A_{2(j)}^{~(n)} \frac{\Delta\tau}{\Delta\eta}
       \bigg] \varepsilon_{q(j-1)}^{~(n)} \nonumber \\
&& \hspace*{12mm}
   -\frac{1}{2} \bigg[  A_{3(j)}^{~(n)}-A_{4(j)}^{~(n)} \frac{\Delta\tau}{\Delta\eta}
   \bigg]\alpha_{q(j+1)}^{~(n)}
   -\frac{1}{2} \bigg[ A_{3(j)}^{~(n)}+A_{4(j)}^{~(n)} \frac{\Delta\tau}{\Delta\eta}
   \bigg]\alpha_{q(j-1)}^{~(n)} =0
\end{eqnarray}
and
\begin{eqnarray}
   && c^{2~(n+1)}_{s~(j)}   \bigg[ B_{1(j)}^{~(n)} \bigg]\varepsilon_{q(j)}^{~(n+1)}
     +\bigg[ B_{3(j)}^{~(n)} \bigg] \alpha_{q(j)}^{~(n+1)} \hspace*{75mm}
\nonumber \\
   && \hspace*{12mm}
     -\frac{c^{2(n)}_{s(j+1)} }{2} \bigg[B_{1(j)}^{~(n)} - B_{2(j)}^{~(n)} \frac{\Delta\tau}{\Delta\eta}
       \bigg] \varepsilon_{q(j+1)}^{~(n)}
     -\frac{ c^{2(n)}_{s(j-1)} }{2} \bigg[ B_{1(j)}^{~(n)} + B_{2(j)}^{~(n)} \frac{\Delta\tau}{\Delta\eta}
       \bigg] \varepsilon_{q(j-1)}^{~(n)} \nonumber \\
&& \hspace*{12mm}
   -~\frac{\quad1\quad}{2}\bigg[  B_{3(j)}^{~(n)}-B_{4(j)}^{~(n)} \frac{\Delta\tau}{\Delta\eta}
   \bigg]\alpha_{q(j+1)}^{~(n)}
   -~\frac{\quad1\quad}{2} \bigg[ B_{3(j)}^{~(n)}+B_{4(j)}^{~(n)} \frac{\Delta\tau}{\Delta\eta}
   \bigg]\alpha_{q(j-1)}^{~(n)} =0 .
\end{eqnarray}
Eliminating $\alpha_{q(j)}^{~(n+1)}$ from the above two equations,
one obtains
\begin{eqnarray}
  F_{10(j)}^{~~(n)} ~\varepsilon_{q(j)}^{~(n+1)} +
  F_{1+(j)}^{~~(n)} ~\varepsilon_{q(j+1)}^{~(n)} +
  F_{0-(j)}^{~~(n)} ~\varepsilon_{q(j-1)}^{~(n)} +
  G_{0+(j)}^{~~(n)} ~\alpha_{q(j+1)}^{~(n)}      +
  G_{0-(j)}^{~~(n)} ~\alpha_{q(j-1)}^{~(n)}   = 0 ,
  \label{eq:A07}
\end{eqnarray}
where
\begin{eqnarray*}
&& F_{10(j)}^{~~(n)} \equiv 1- c_{s(j)}^{2(n+1)} ~[v_{q(j)}^{~(n)}]^2  , \\
&& F_{0+(j)}^{~~(n)} \equiv -\frac{1}{2}\left(   1- c_{s(j+1)}^{2(n)} ~[v_{q(j)}^{~(n)}]^2    \right)
    +\frac{1}{2}\left(     1-  c_{s(j+1)}^{2(n)}  \right) \frac{[v_{q(j)}^{~(n)}]}{\tau_n}\frac{\Delta\tau}{\Delta\eta} ,\\
&& F_{0-(j)}^{~~(n)} \equiv  -\frac{1}{2} \left(   1- c_{s(j-1)}^{2(n)} ~[v_{q(j)}^{~(n)}]^2    \right)
    -\frac{1}{2} \left(     1-  c_{s(j-1)}^{2(n)}  \right) \frac{[v_{q(j)}^{~(n)}]}{\tau_n}\frac{\Delta\tau}{\Delta\eta}, \\
&& G_{0+(j)}^{~~(n)} \equiv +\frac{1}{2} (1-[v_{q(j)}^{~(n)}]^2) (~\varepsilon_{q(j)}^{~(n)}+P_{q(j)}^{~(n)} )
     \frac{1}{\tau_n} \frac{\Delta\tau}{\Delta\eta} ,\\
&& G_{0-(j)}^{~~(n)} \equiv -\frac{1}{2} (1-[v_{q(j)}^{~(n)}]^2) (~\varepsilon_{q(j)}^{~(n)}+P_{q(j)}^{~(n)} )
     \frac{1}{\tau_n} \frac{\Delta\tau}{\Delta\eta} .
\end{eqnarray*}
One can now find $\varepsilon_{q(j)}^{~(n+1)}$ by solving the
non-linear Eq.~(\ref{eq:A07}). For $v_{q(j)}^{~(n)}=0$ (i.e., for
the scaling case where $\eta=\alpha$) one obtains
\begin{eqnarray}
  \varepsilon_{q(j)}^{~(n+1)} - \varepsilon_{q(j)}^{~(n)}
+\frac{\varepsilon_{q(j)}^{(n)}+P_{q(j)}^{(n)}}{\tau_n}    \Delta\tau =0
\end{eqnarray}
where relations $\alpha_{q(j+1)}^{~(n)} -
\alpha_{q(j-1)}^{~(n)}=2\Delta\eta$ and
$\frac{1}{2}[\varepsilon_{q(j+1)}^{~(n)}+\varepsilon_{q(j-1)}^{~(n)}]=
\varepsilon_{q(j)}^{~(n)}$ were used. After finding
$\varepsilon_{q(j)}^{~(n+1)} $ one can find $\alpha_{q(j)}^{~(n+1)}
$ by using the following recurrence formula:
\begin{eqnarray}
   \alpha_{q(j)}^{~(n+1)} \!\!\!&=&\!\!\! \frac{1}{2}\left[\alpha_{q(j+1)}^{~(n)} +\alpha_{q(j-1)}^{~(n)}  \right]
                               - \frac{1}{2}\left[\alpha_{q(j+1)}^{~(n)} -\alpha_{q(j-1)}^{~(n)}  \right]
                                  \frac{v_{q(j)}^{~(n)}}{\tau_n} \frac{\Delta\tau}{\Delta\eta} \nonumber \\
 &&+ \frac{v_{q(j)}}{2} \left[
    \frac{ c_{s(j+1)}^{2(n)} }{1+c_{s(j)}^{2(n)}  } \frac{ \varepsilon_{q(j+1)}^{(n)}}{ \varepsilon_{q(j)}^{(n)} }  +
    \frac{ c_{s(j-1)}^{2(n)} }{1+c_{s(j)}^{2(n)}  } \frac{ \varepsilon_{q(j-1)}^{(n)}}{ \varepsilon_{q(j)}^{(n)} }  -
2~ \frac{ c_{s(j)}^{2(n+1)} }{1+c_{s(j)}^{2(n)}  } \frac{ \varepsilon_{q(j)}^{(n+1)}}{ \varepsilon_{q(j)}^{(n)} } \right]  \nonumber \\
&& - ~\frac{~1~}{2}~ \left[
 \frac{ c_{s(j+1)}^{2(n)} }{1+c_{s(j)}^{2(n)}  } \frac{ \varepsilon_{q(j+1)}^{(n)}}{ \varepsilon_{q(j)}^{(n)} }  -
    \frac{ c_{s(j-1)}^{2(n)} }{1+c_{s(j)}^{2(n)}  } \frac{ \varepsilon_{q(j-1)}^{(n)}}{ \varepsilon_{q(j)}^{(n)} }
\right] \frac{1}{\tau_n}\frac{\Delta\tau}{\Delta\eta}~.
\end{eqnarray}
\end{widetext}

\section{\label{sec:no-approx}Inadequacy of the simple expansion in $|q-1|$}

From the previous experience in applying $q$-statistics to
multiparticle production \cite{E1a,E1b,E2,Kq,QlQt,E3} we know that
$|q-1| <1$. It seems then natural to argue that (see, for example,
\cite{E1b,S0}) one could simply expand $f_q(x,p) = \left[
1-(1-q)\frac{p^{\mu}u_{\mu}}{k_B T(x)}\right]^{1/(1-q)}$ from
Eq.~(\ref{eq:fq}) in $z= 1-q$, retaining only terms linear in $z$ and
get:
\begin{eqnarray}
f_q(x,p) \!\! &=&\!\! f(z) = \left[ 1 - z\cdot A
\right]^{\frac{1}{z}}
\!\! \equiv \!\! \left[ \frac{1}{z}\ln (1 - z\cdot A ) \right] \nonumber \\
\!\! &\simeq&\!\! f(z=0) + z\cdot \frac{d f(z)}{dz}\Big|_{z=0}
\end{eqnarray}
(here $A = A(x,p) = \frac{p^{\mu}u_{\mu}(x)}{k_BT}$ and the
arguments $(x,p)$ are suppressed for clarity).

However, such expansion can only be performed under some
conditions, which we shall clarify in what follows. It is
straightforward to show that to gets first step of expansion,
\begin{eqnarray}
f(z) &\simeq& \exp \left[ -A \cdot \left( 1 + \frac{A}{2}\cdot z +
\dots \right)\right] \nonumber \\
&& = \exp \left[ -A\right]\cdot \exp\left[
-\frac{A^2}{2}\cdot z - \dots \right], \label{eq:Iz2order}
\end{eqnarray}
it is necessary that
\begin{equation}
z\cdot A(x,p) < 1  .\label{eq:firstcond}
\end{equation}
The second step needed is to additionally expand the exponent and
this requires
\begin{equation}
z\cdot A^2(x,p) < 2  .\label{eq:secondcond}
\end{equation}
When this is satisfied, one finally gets $f_q(x,p)$ in terms of
$f_{q=1}(x,p)$ only,
\begin{equation}
f_q(x,p) \simeq f_{q=1}(x,p) + (1-q) \left[ 1 - \frac{A^2(x,p)}{2}
\right] f_{q=1}(x,p). \label{eq:final}
\end{equation}

At first this procedure looks very promising because using it one
gets
\begin{eqnarray}
{\cal T}_{q}^{\mu\nu}\!\!\!&\equiv&\!\!\! {\cal T}_{q=1}^{\mu\nu}
 +(q-1) {\tau}_q^{\mu\nu} ,\label{eq:Tqq}
\end{eqnarray}
where ${\cal T}_{q=1}$ is the usual energy-momentum tensor for the
equilibrium of the Boltzmann-Gibbs statistics, i.e. the one
usually used when describing an ideal fluid,
\begin{eqnarray}
  {\cal T}_{q=1}^{\mu\nu}\equiv
   \frac{g}{(2\pi)^3}\int\!\!
   \frac{d^3p}{p^0} ~p^{\mu}  p^{\nu} \exp\left(-\frac{p\cdot
   u}{T}\right), \label{eq:T0}
\end{eqnarray}
whereas the nonextensive correction tensor $\tau^{\mu\nu}_q$ is given
by
\begin{eqnarray}
  && \tau_q^{\mu\nu} \equiv
  \frac{g}{(2\pi)^3}
  \!\!\int\!\!\frac{d^3p}{p^0}~p^{\mu}p^{\nu} \cdot \nonumber \\
  && \hspace*{5mm}\cdot \exp\left(-\frac{p\cdot u}{T}\right)
   \left[
   -\left(\frac{p\cdot u}{T}\right)
  +\frac{1}{2}\left(\frac{p\cdot u}{T}\right)^2
   \right]. \label{eq:Pi0}
\end{eqnarray}
However, in our case condition (\ref{eq:secondcond}) would impose
too severe constraints on the allowed $q$ and the region of phase
space, $p$ and $x$, considered rendering this approximation rather
unpractical for our purposes.

\section{\label{sec:shear_visco} Derivation of Eq.~(\ref{eq:sigma_full1}) and $\eta/s$, Eq.~(\ref{eq:sigma_full2}).}

The entropy production is given by
\begin{eqnarray}
\sigma^{\mu}_{{\rm full} ;\mu} \!\!&=&\!\! \left[ \xi \Psi^{\mu}  \right]_{;\mu}
= \frac{\partial_{\mu}\xi}{\xi} \xi\Psi^{\mu} + \xi \Psi^{\mu}_{;\mu}
\end{eqnarray}
where
\begin{eqnarray}
\Psi^{\mu} \!\!&=&\!\! -\frac{W^{\nu}W_{\nu}}{3\Pi T} u^{\mu}+ \frac{ W_{\nu}}{2\Pi T} \pi^{\mu\nu} \nonumber \\
\!\!&=&\!\! \frac{-1}{6\Pi T}  \left\{ 2W^{\nu}W_{\nu}u^{\mu} -3W_{\nu} \pi^{\mu\nu} \right\} .
\end{eqnarray}
Then,
\begin{eqnarray}
 \Psi^{\mu}_{;\mu} \!\!&=&\!\!  \left( \frac{\partial_{\mu} \Pi}{\Pi} + \frac{\partial_{\mu} T}{T} \right) ~ \Psi^{\mu}
+ \frac{-1}{6\Pi T}  \psi^{\mu}_{;\mu},
\end{eqnarray}
where $\psi^{\mu} \equiv 2W^{\nu}W_{\nu}u^{\mu} -3W_{\nu} \pi^{\mu\nu}$.
The 
$\psi^{\mu}_{;\mu}$ is explicitly written as
\begin{eqnarray}
\psi^{\mu}_{;\mu} \!\!&=&\!\! \left\{ 2W^{\nu}W_{\nu}u^{\mu} -3W_{\nu} \pi^{\mu\nu} \right\}_{;\mu} \hspace*{25mm}\nonumber \\
\!\!&=&\!\!
\left[ 4 W_{\nu;\mu}u^{\mu} +2W_{\nu}u^{\mu}_{;\mu}-3\pi^{\mu}_{\nu;\mu} \right] W^{\nu}
+\left[ -3W_{\nu;\mu} \right]\pi^{\mu\nu} \nonumber
\end{eqnarray}
and
\begin{eqnarray}
&&\frac{-1}{6\Pi T}\psi^{\mu}_{;\mu}
=\nonumber \\
&& \hspace*{5mm}\frac{-1}{\Pi}\left[ \frac{2}{3} W_{\nu;\mu}u^{\mu} +\frac{1}{3}W_{\nu}u^{\mu}_{;\mu} -\frac{1}{2}\pi^{\mu}_{\nu;\mu} \right] \frac{W^{\nu}}{T}
\nonumber \\
&&\hspace*{5mm} + \frac{-1}{\Pi}\left[ -\frac{1}{2}W_{\nu;\mu}\right] \frac{\pi^{\mu\nu}}{T}.
\end{eqnarray}
Hence we obtain that
\begin{eqnarray}
\xi \Psi^{\mu}_{;\mu} \!\!&=&\!\! \xi  \left( \frac{\partial_{\mu} \Pi}{\Pi} + \frac{\partial_{\mu} T}{T}
 \right) \Psi^{\mu}  \nonumber \\
&& \hspace*{5mm} -\frac{\xi}{\Pi}\left[ \frac{2}{3} W_{\nu;\mu} u^{\mu}+\frac{1}{3}W_{\nu}u^{\mu}_{;\mu} -\frac{1}{2}\pi^{\mu}_{\nu;\mu} \right] \frac{W^{\nu}}{T}
\nonumber \\
&&\hspace*{5mm} - \frac{\xi}{\Pi}\left[ -\frac{1}{2}W_{\nu;\mu}\right] \frac{\pi^{\mu\nu}}{T} \nonumber \\
&&\hspace*{-5mm} = \xi  \left( \frac{\partial_{\mu} \Pi}{\Pi} + \frac{\partial_{\mu} T}{T}
 \right) \Psi^{\mu}  -Y_{\nu}\frac{W^{\nu}}{T} + Z_{\mu\nu} \frac{\pi^{\mu\nu}}{T}
\end{eqnarray}
Finally one arrives at
\begin{eqnarray}
\sigma^{\mu}_{{\rm full} ;\mu} \!\!&=&\!\! \left[ \xi \Psi^{\mu}  \right]_{;\mu} \nonumber \\
\!\!&=&\!\!
\frac{\xi}{\Pi}  \left( \frac{\partial_{\mu}\xi}{\xi}+
\frac{\partial_{\mu} \Pi}{\Pi} + \frac{\partial_{\mu} T}{T}
 \right) \Pi \Psi^{\mu}
 \nonumber \\ && \hspace*{15mm}
-Y_{\nu}\frac{W^{\nu}}{T} + Z_{\mu\nu} \frac{\pi^{\mu\nu}}{T}
\nonumber \\
\!\!&=&\!\! -X_{\mu} T \Psi^{\mu}\frac{\Pi}{T} -Y_{\nu}\frac{W^{\nu}}{T} + Z_{\mu\nu} \frac{\pi^{\mu\nu}}{T} \nonumber \\
\!\!&=&\!\! -X_{\mu}  ( wu^{\mu} ) \frac{\Pi}{T}
 -\tilde{Y}_{\nu}  \frac{W^{\nu}}{T}  + Z_{\mu\nu} \frac{\pi^{\mu\nu}}{T}.
\end{eqnarray}
This is Eq.~(\ref{eq:sigma_full1}). Using now
$\pi^{\mu\nu}W_{\nu}=-2\Pi W^{\mu}$ one gets that
\begin{eqnarray}
\sigma^{\mu}_{{\rm full} ;\mu}  \!\!&=&\!\!  -X_{\mu}  ( wu^{\mu} ) \frac{\Pi}{T}
 -\tilde{Y}_{\nu}  \frac{W^{\nu}}{T}  + Z_{\mu\nu} \frac{\pi^{\mu\nu}}{T} \nonumber \\
 \!\!&=&\!\!
 -X_{\mu}  ( wu^{\mu} ) \frac{\Pi}{T}
+\left[  \frac{ \tilde{Y}_{\nu}    W_{\nu}}{2\Pi T}
+ Z_{\mu\nu} \right]  \frac{\pi^{\mu\nu}}{T} \nonumber \\
 \!\!&=&\!\!
 -X_{\mu}  ( wu^{\mu} ) \frac{\Pi}{T} +\tilde{Z}_{\mu\nu}
 \frac{\pi^{\mu\nu}}{T},
\label{eq:ApdxD7}
\end{eqnarray}
which is Eq. (\ref{eq:sigma_full2}).

\end{document}